\documentclass[11pt]{article}
\usepackage[normalem]{ulem}
\usepackage{float,eurosym}
\usepackage{xspace}
\usepackage[latin1]{inputenc}
\usepackage[usenames,dvipsnames]{color}
\usepackage{amssymb}
\usepackage{amsmath}
\usepackage{amsthm}
\usepackage{enumerate}
\usepackage{graphicx}
\usepackage{caption}
\usepackage{subcaption}
\usepackage{multirow}

\newtheorem{Proposition}{Proposition}[part]

\newtheorem{Lemma}{Lemma}[part]

\newtheorem{Remark}{Remark}[part]
\newtheorem{Example}{Example}[part]
\def\bl{\begin{Lemma}}
\def\el{\end{Lemma}}

\newcommand{\e}{\textrm{e}}

\def \Frac{\displaystyle\frac}

\def \R{\mathbb{R}}
\def \M{\mathbb{M}}

\def \E{\mathbb{E}}
\def \Fc{\mathbb{F}}

\def \P{\mathbb{P}}
\def \Q{\mathbb{Q}}

\def \S{\mathbb{S}}

\def \Fc{{\cal F}}

\def \1{{\mathds 1}}

\def \1b{{\bf 1}}

\def\beqs{\begin{eqnarray*}}
\def\enqs{\end{eqnarray*}}
\def\beq{\begin{eqnarray}}
\def\enq{\end{eqnarray}}

\newcommand{\nc}{\newcommand}
\nc{\esssup}{\mathop{\mathrm{ess\;sup}}}

\usepackage{chicago}
\usepackage{graphicx}
\usepackage{epstopdf}
 \def\s#1{\begin{color}[rgb]{0.00,0.00,1.00}{#1}\end{color}}\def\s#1{#1}
\newcommand{\PR}{\mathbb{P}}
\def\cite{\citeN}
\def\sp{,\,}
\def\thea{\alpha}\def\thea{{c_{1}}}
\def\theb{\beta}\def\theb{b_{1}}
\def\a{\alpha}\def\a{{c_{i}}}
\def\b{\beta}\def\b{{b_{i}}}

\def\da{\dot{\a}} \def\da{\dot{c}_{i}} 
\def\db{\dot{\b}}\def\db{\dot{b}_{i}} 

\def\theC{C}\def\theC{A^{(1)}}
\def\S{S}
\def\C{\C}\def\C{A^{(1)}}\def\C{A^{(i)}}

\def\thep{p}\def\thep{h}
\def\theq{q}\def\theq{k}
\def\tvartheta{\tilde{\vartheta}}
\def\rd{d}
\newcommand{\bea}{\begin{eqnarray*}}
\newcommand{\eea}{\end{eqnarray*}}
\newcommand{\beqa}{\begin{eqnarray}}
\newcommand{\eeqa}{\end{eqnarray}}
\def\bal{\begin{aligned}}
\def\eal{\end{aligned}}
\newcommand{\beql}[1]{\beqa\label{#1}\bal}
\newcommand{\eeql}{\eal\eeqa}
\newcommand{\bel}{\bea\bal}
\newcommand{\eel}{\eal\eea}
\def\qqq{\quad\quad\quad}
\def\Finproof{\rule{4pt}{6pt}}

\def\EQ{\mathbb{Q}}\def\EQ{\mathbb{E}^\mathbb{Q}}
\def\ER{\mathbb{R}}\def\ER{\mathbb{E}^\mathbb{R}}
\def\EM{\mathbb{M}}\def\EM{\mathbb{E}^\mathbb{M}}

\def\notTheta{\Theta}
\def\tTheta{\widetilde\Theta}
\def\Ts{\bar{T}}\def\Ts{\overline{T}}\def\Ts{T}
\def\mygb{\bar{g}}\def\mygb{\tilde{g}}\def\mygb{f}
\def\mygh{\tilde{f}}
\def\ttt{t \in [0,\Ts]}
\def\bll#1{
\beqa\label{#1}\bal
}\def\lel{\eal\eeqa}
\def\thisR{\thisR }
\def\thisRm{\overline{\thisR}}
\def\thisR{R}\def\thisR{\thisR }\def\thisR{R_b}
\def\thisRm{{\bar{R}}}\def\thisRm{{\overline{\thisR}}}
\def\thisRm{R_c}

\def\Gp{\Gamma^+}
\def\Gm{\Gamma^-}

\def\thebm{b^-}\def\thebm{\bar{b}}
\def\thelambda{\lambda^+}\def\thelambda{\lambda}
\def\thelambdam{\lambda}\def\thelambdam{\lambda^-}\def\thelambdam{\bar{\lambda}}
\def\thenewb{b}
\def\Rf{\mathfrak{R}}\def\Rf{\mathfrak{r}}\def\Rf{\mathfrak{r}^b}\def\Rf{\bar{R}_b}

\def\Nd{1}\def\Nd{N\delta}
\def\theX{X}
\def\Bd{B}\def\Bd{P}

\def\qr#1{\eqref{#1}}
\def\sr#1{Section~\ref{#1}}
\newcommand{\ebe}{\begin{enumerate}[1)]\setlength{\baselineskip}{13pt}\setlength{\parskip}{5pt}}
\newcommand{\dbe}{\end{enumerate}}
\begin{document}
\title{Rational Multi-Curve Models with \\ Counterparty-Risk Valuation Adjustments
}
\author{St\'ephane Cr\'epey $ ^{1},$ Andrea Macrina $ ^{2,3}$,
Tuyet Mai Nguyen$ ^{1}$, David Skovmand $^4$
\\\\
$ ^1 $ Laboratoire de Math\'ematiques et Mod\'elisation d'\'Evry, France\vspace{4pt}\\
$ ^2 $ Department of Mathematics, University College London, United Kingdom\vspace{4pt}\\$ ^3 $ Department of Actuarial Science, University of Cape Town, South Africa\vspace{6pt}\\
$^4$ Department of Finance, Copenhagen Business School, Denmark
}
\date{\today}
\maketitle
\vspace{-1cm}
\begin{abstract}
We develop a multi-curve term structure setup in which the modelling ingredients are expressed by rational functionals of Markov processes.  We calibrate to LIBOR swaptions data and show that a rational two-factor lognormal multi-curve model is sufficient to match market data with accuracy. 
We elucidate the relationship between the models developed and calibrated under a risk-neutral measure $\Q$ and their consistent equivalence class under the real-world probability measure $\PR$. The consistent $\PR$-pricing models are applied to compute the risk exposures which may be required to comply with regulatory obligations. In order to compute counterparty-risk valuation adjustments, such as CVA, we show how positive default intensity processes with rational form can be derived. We flesh out our study by applying the results to a basis swap contract.
\\ \vspace{-0.2cm}
\end{abstract}

\section{Introduction}\label{s:intro}
In this work we endeavour to develop multi-curve interest rate models which extend to counterparty risk models in a consistent fashion. The aim is the pricing and risk management of financial instruments with price models capable of discounting at multiple rates (e.g. OIS and LIBOR) and which allow for corrections in the asset's valuation scheme so to adjust for counterparty-risk inclusive of credit, debt, and liquidity risk. We thus propose factor-models for (i) the Overnight Index Swap (OIS) rate, (ii) the London Interbank Offer Rate (LIBOR), and (iii) the default intensities of two counterparties involved in bilateral OTC derivative transactions. The three ingredients are characterised by a feature they share in common: the rate and intensity models are all rational functions of the underlying factor processes. In choosing this class of models, we look at a number of properties we would like the models to exhibit. They should be flexible enough to allow for the pricing of a range of financial assets given that all need discounting and no security is insulated from counterparty-risk. Since we have in mind the pricing of assets as well as the management of risk exposures, we also need to work within a setup that maintains price consistency under various probability measures. We will for instance want to price derivatives by making use of a risk-neutral measure $\mathbb{Q}$ while analysing the statistics of risk exposures under the real-world measure $\mathbb{P}$. This point is particularly important when we calibrate the interest rate models to derivatives data, such as (implied) volatilities, and then apply the calibrated models to compute counterparty-risk valuation adjustments to comply with regulatory requirements. The presented rational models allow us to develop a comprehensive framework that begins with an OIS model, evolves to an approach for constructing the LIBOR process, includes the pricing of fixed-income assets and model calibration, analyses risk exposures, and concludes with a credit risk model that leads to the analysis of counterparty-risk valuation adjustments (XVA).  

The issue of how to model multi-curve interest rates and incorporate counterparty-risk valuation adjustments in a pricing framework has motivated much research. For instance, research on multi-curve interest rate modelling is presented in  \citeN{KijimaTanakaWong09}, \citeN{Kenyon10}, \citeANP{Henrard07} (\citeyearNP{Henrard07}, \citeyearNP{Henrard09}, \citeyearNP{Henrard14}), \citeN{Bianchetti10}, Mercurio (\citeyearNP{Mercurio2010}, \citeyearNP{Mercurio2010a}, \citeyearNP{merc}), \citeANP{Fujii2009} (\citeyearNP{Fujii2009}, \citeyearNP{Fujii2010}), \citeN{MoreniPallavicini10}, \citeN{BianchettiMorini13}, \citeN{FilipovicTrolle11} and \cite{CrepeyGrbacNgorSkovmand13}. On counterparty-risk valuation adjustment, we mention two recent books by \citeN{BrigoMoriniPallavicini12} and \citeN{CrepeyBieleckiBrigo14}; more references are given as we go along. Pricing models with rational form have also appeared before. \cite{fh} pioneered such pricing models and in particular introduced the so-called rational log-normal model for discount bond prices. For further contributions and studies in this context we refer to \cite{Rut}, \cite{ds} and \cite{HuntKennedy}. More recent work on rational pricing models include \cite{BH}, \cite{hr}, \cite{BHMackie}, \cite{ahtt}, \cite{FLT},  \cite{AMPAP}, and \cite{NguyenSeifried2014}. However, as far as we know, the present paper is the first to apply such models in a multi-curve setup, along with \cite{NguyenSeifried2014}, who develop a rational multi-curve model based on a multiplicative spread. It is the only one to deal with XVA computations. We shall see that, despite the simplicity of these models, their performances in these regards are comparable to those by \citeN{CrepeyGrbacNgorSkovmand13} or \citeANP{MoreniPallavicini10} (\citeyearNP{MoreniPallavicini13}, \citeyearNP{MoreniPallavicini10}). Other recent related research includes \cite{FLT}, for the study of unspanned volatility and its regulatory implications, \cite{CuchieroKeller-ResselTeichmann12}, for moment computations in financial applications, and \cite{ChengTehranchi}, motivated by stochastic volatility modelling.  

We give a brief overview of this paper. In Section \ref{s:pkba}, we introduce the rational models for multi-curve term structures whereby we derive the forward LIBOR process by pricing a forward rate agreement under the real-world probability measure. In doing so we apply a pricing kernel model. The short rate model arising from the pricing kernel process is then assumed to be a proxy model for the OIS rate. In view of derivative pricing in subsequent sections, we also derive the multi-curve interest rate models by starting with the risk-neutral measure. We call this method ``bottom-up risk-neutral approach".  In Section \ref{s:pg}, we perform the so-called ``clean valuation" of swaptions written on LIBOR, and analyse three different specifications for the OIS-LIBOR dynamics. We explain the advantages one gains from the chosen ``codebook" for the LIBOR process, which we model as a rational function where the denominator is in fact the stochastic discount factor associated with the utilised probability measure. In Section \ref{s:cal}, we calibrate the three specified multi-curve models and assess them for the quality of fit and on positivity of rates and spread. We conclude by singling out a \s{two-factor} lognormal OIS-LIBOR model for its satisfactory calibration properties and acceptable level of tractability. In Section \ref{s:expos}, we price a basis swap in closed form without taking into account counterparty-risk, that is we again perform a ``clean valuation". In this section we take the opportunity to show the explicit relationship in our setup between pricing under an equivalent measure and the real-world measure. We compute the risk exposure associated with holding a basis swap and plot the quantiles under both probability measures for comparison. As an example, we apply L\'evy random bridges to describe the dynamics of the factor processes under $\mathbb{P}$. This enables us to interpret the re-weighting of the risk exposure under $\mathbb{P}$ as an effect that could be related to, e.g., ``forward guidance'' provided by a central bank. In the last section, we present default intensity processes with rational form and compute XVA, that is, the valuation adjustments due to credit, debt, and liquidity risk.
\section{Rational multi-curve term structures}\label{s:pkba}
We model a financial market by a filtered probability space $(\Omega,\Fc,\P,\{\Fc_t\}_{0\leq t})$, where $\P$ denotes the real probability measure and $\{\Fc_t\}_{0\leq t}$ is the market filtration. The no-arbitrage pricing formula for a generic (non-dividend-paying) financial asset with price process $\{S_{tT}\}_{0\leq t\leq T}$, which is characterised by a cash flow $S_{TT}$ at the fixed date $T$,  is given by
\begin{equation}\label{pf-formula}
S_{tT} = \dfrac{1}{\pi_t}\E^{\P}[\pi_T S_{TT}\,\vert\,\Fc_t],
\end{equation}
where $\{\pi_t\}_{0\leq t\leq U}$ is the pricing kernel embodying the inter-temporal discounting and risk-adjustments, see e.g. \cite{HuntKennedy}. Once the model for the pricing kernel is specified, the OIS discount bond price process $\{\Bd_{tT}\}_{0\leq t\leq T\leq U}$ is determined as a special case of formula (\ref{pf-formula}) by
\begin{equation}\label{discountbond}
\Bd_{tT} = \dfrac{1}{\pi_t}\E^{\P}[\pi_T\,\vert\,\Fc_t] .
\end{equation}
The associated OIS short rate of interest is obtained by
\begin{equation}\label{shortrate}
r_t=-\left(\partial_T \ln \Bd_{tT}\right)\vert_{T=t},
\end{equation}
where it is assumed that the discount bond system is differentiable in its maturity parameter $T$. \s{The rate $\{r_t\}$ is non-negative if the pricing kernel $\{\pi_t\}$ is a supermartingale and vice versa.} We next go on to infer a pricing formula for financial derivatives written on LIBOR. In doing so, we also derive a price process (\ref{LIBOR-proc}) that we identify as determining the dynamics of the forward LIBOR or, as we shall call it, the LIBOR process. It is this formula for the LIBOR process that reveals the nature of the so-called multi-curve term structure whereby the OIS rate and the LIBOR rates of different tenors are treated as distinct discount rates.
\subsection{Generic multi-curve interest rate models}
We derive multi-curve pricing models for securities written on the LIBOR by starting with the valuation of a forward rate agreement (FRA). We consider $0\le t\le T_{0}\le T_{2}\le\cdots\le T_i\le\cdots\le T_n$, where $T_0 ,T_i,\ldots,T_n$ are fixed dates, and let $N$ be a notional, $K$ a strike rate and $\delta_i=T_i-T_{i-1}$. The fixed leg of the FRA contract is given by $NK\delta_i$ and the floating leg payable in arrear at time $T_i$ is modelled by  $N\delta_i L(T_i;T_{i-1},T_i),$ where the random rate $L(T_i;T_{i-1},T_i)$ is $\Fc_{T_{i-1}}$-measurable. Then we define the net cash flow at the maturity date $T_i$ of the FRA contract to be
\begin{equation}
H_{T_i}=N\delta_i\left[K-L(T_i;T_{i-1},T_i)\right].
\end{equation}
The FRA price process is then given by an application of \eqref{pf-formula}, that is, for $0\leq t\leq T_{i-1},$ by 
\begin{eqnarray}\nonumber
H_{tT_i}&=&\Frac{1}{\pi_t}\E^{\PR}\left[\pi_{T_i} H_{T_i}\,\big\vert\,\Fc_t\right]\\
		&=&N\delta_i\left[K\Bd_{tT_i}-L(t,T_{i-1},T_i)\right],\label{FRApp}
\end{eqnarray}
where we define the (forward) LIBOR process by
\begin{equation}\label{LIBOR-proc}
L(t;T_{i-1},T_i):=\Frac{1}{\pi_t}\E^{\PR}\left[\pi_{T_i}L(T_i;T_{i-1},T_i)\,\big\vert\,\Fc_t\right].
\end{equation}
The fair spread of the FRA at time $t$ (the value $K$ at time $t$ such that $H_{tT_i}=0$) is then expressed in terms of $L(t;T_{i-1},T_i)$ by
\beql{fra-proc}
K_t = \Frac {L(t;T_{i-1},T_i)} {\Bd_{t T_i}}.
\eeql
For times up to and including $T_{i-1},$ our LIBOR process can be written in terms of a conditional expectation of an ${\cal F}_{T_{i-1}}$-measurable random variable. In fact, for $t\leq T_{i-1},$
\begin{eqnarray}
\E^{\PR}\left[\pi_{T_i} L(T_i;T_{i-1},T_i) \,\big\vert\,\Fc_t\right]&=&\E^{\PR}\left[\E^{\PR}\left[\pi_{T_i}L(T_i;T_{i-1},T_i)\,\big\vert\,\Fc_{T_{i-1}}\right]\,\big\vert\,\Fc_t\right]\\
&=&\E^{\PR}\left[\E^{\PR}\left[\pi_{T_i}\,\big\vert\,\Fc_{T_{i-1}}\right]L(T_i;T_{i-1},T_i)\,\big\vert\,\Fc_t\right],
\end{eqnarray}
and thus
\begin{equation}\label{LIBOR-Model}
L(t,T_{i-1},T_i) =\Frac{1}{\pi_t}\E^{\PR}\left[\E^{\PR}\left[\pi_{T_i}\,\big\vert\,\Fc_{T_{i-1}}\right] 
L(T_i;T_{i-1},T_i)\,\big\vert\,\Fc_t\right].
\end{equation}

The (pre-crisis) classical approach to LIBOR modelling defines the price process $\{H_{tT_i}\}$ of a FRA by
\begin{equation}
H_{tT_i}=N\left[(1+\delta_i K)\Bd_{tT_i}-\Bd_{tT_{i-1}}\right],
\end{equation}
see, e.g., \cite{HuntKennedy}. By equating with (\ref{FRApp}), we see that the classical single-curve LIBOR model is obtained in the special case where
\begin{equation}
L(t;T_{i-1},T_i)=\Frac{1}{\delta_i}\left(\Bd_{tT_{i-1}}-\Bd_{tT_i}\right).
\end{equation}

\begin{Remark}\label{spread-pos}\em
In normal market conditions, one expects the positive-spread relation \linebreak $L(t;T,T+\delta_i)<L(t;T,T+\delta_j)$, for tenors $\delta_j>\delta_i$, to hold. We will return to this relationship in Section \ref{s:cal} where various model specifications are calibrated and the positivity of the spread is checked. LIBOR tenor spreads play a role in the pricing of basis swaps, which are contracts that exchange LIBOR with one tenor for LIBOR with another, different tenor (see Section \ref{s:expos}).  For recent work on multi-curve modelling with focus on spread modelling, we refer to \cite{CuchieroFontanaGnoatto14}. 
\end{Remark}
\subsection{Multi-curve models with rational form}\label{ss:multi}
In order to construct explicit LIBOR processes, the pricing kernel $\{\pi_t\}$ and the random variable $L(T_i;T_{i-1},T_i)$ need to be specified in the definition (\ref{LIBOR-proc}). For reasons that will become apparent as we move forward in this paper, we opt to apply the rational pricing models proposed in \cite{Mac}. These models bestow a rational form on the price processes, here intended as a ``quotient of summands'' (slightly abusing the terminology that usually refers to a ``quotient of polynomials''). This explains the terminology in this paper when referring to the class of multi-curve term structures, or to generic asset price models, and later also to the models for counterparty risk valuation adjustments.  

The basic pricing model with rational form for a generic financial asset (for short ``rational pricing model'') that we consider is given by  
\begin{equation}\label{S-proc}
S_{tT} = \dfrac{S_{0T} + b_2(T)A_t^{(2)}+ b_3(T)A_t^{(3)}}{\Bd_{0t} + b_1(t)A_t^{(1)}},
\end{equation}
where $S_{0T}$ is the value of the asset at $t=0$. There may be more $bA$-terms in the numerator, but two (at most) will be enough for all our purposes in this work.
For $0\le t\le T$ and $i=1,2,3$, $b_i(t)$ are deterministic functions and $A^{(i)}_t = A_i(t,X_t^{(i)})$ are martingale processes, not necessarily under $\PR$ but under an equivalent martingale measure $\M$, which are driven by $\M$-Markov processes $\{X^{(i)}_t\}$. The details of how the expression (\ref{S-proc}) is derived from the formula (\ref{pf-formula}), and in particular how explicit examples for  $\{A^{(i)}_t\}$ can be constructed, are shown in \cite{Mac}. Here we only give the pricing kernel model associated with the price process (\ref{S-proc}), that is
\begin{equation}
\pi_t=\frac{\pi_0}{M_0}\left[\Bd_{0t}+b_1(t)A^{(1)}_t\right]M_t,
\end{equation}
where $\{M_t\}$ is the $\PR$-martingale that induces the change of measure from $\PR$ to an auxiliary measure $\M$ under which the $\{A^{(i)}_t\}$ are martingales. 
The deterministic functions $\Bd_{0t}$ and $b_1(t)$ are defined such that $\Bd_{0t}+b_1(t)A^{(1)}_t$ is a non-negative $\M$-supermartingale (see e.g. Example \ref{e:theele}), and thus in such a way that $\{\pi_t\}$ is a non-negative $\P$-supermartingale.
By the equations (\ref{discountbond}) and (\ref{shortrate}), it is straightforward to see that 
\begin{equation}\label{bond&rate}
\Bd_{tT} = \dfrac{\Bd_{0T} + b_1(T)A_t^{(1)}}{\Bd_{0t} + b_1(t)A_t^{(1)}},\quad r_t= -\dfrac{\dot{P}_{0t} + \dot{b_1}(t)A_t^{(1)}}{\Bd_{0t} + b_1(t)A_t^{(1)}},
\end{equation}
where the ``dot-notation'' means differentiation with respect to time $t$.

Let us return to the modelling of rational multi-curve term structures and in particular to the definition of the (forward) LIBOR process. Putting equations (\ref{LIBOR-proc}) and (\ref{pf-formula}) in relation, we see that the model (\ref{S-proc}) naturally offers itself as a model for the LIBOR process (\ref{LIBOR-proc}) in the considered setup. Since (\ref{S-proc}) satisfies (\ref{pf-formula}) by construction, so does the LIBOR model
\begin{equation}\label{LIBOR-process}
L(t;T_{i-1},T_i)= \dfrac{L(0;T_{i-1},T_i) + b_2(T_{i-1},T_i)A_t^{(2)}+ b_3(T_{i-1},T_i)A_t^{(3)}}{\Bd_{0t} + b_1(t)A_t^{(1)}}
\end{equation}
satisfy the martingale equation (\ref{LIBOR-proc}) and in particular (\ref{LIBOR-Model}) for $t\le T_{i-1}$. In \cite{Mac} a method based on the use of weighted heat kernels is provided for the explicit construction of the $\M$-martingales $\{A^{(i)}_t\}_{i=1,2}$ and thus in turn for explicit LIBOR processes. The method allows for the development of LIBOR processes, which, if circumstances in financial markets require it, by construction take positive values at all times. 
\subsection{Bottom-up risk-neutral approach}\label{Bottom_up_Risk_neutral}
Since we also deal with counterparty-risk valuation adjustments, we present another scheme for the construction of the LIBOR models, which we call ``bottom-up risk-neutral approach''. As the name suggest, we model the multi-curve term structure by making use of the risk-neutral measure (via the auxiliary measure $\M$) while the connection to the $\PR$-dynamics of prices can be reintroduced at a later stage, which is important for the calculation of risk exposures and their management. ``Bottom-up'' refers to the fact that the short interest rate will be modelled first, then followed by the discount bond price and LIBOR processes. Similarly, in \sr{ss:rcm}, the hazard rate processes for contractual default will be modelled first, and thereafter the price processes of counterparty risky assets will be derived thereof. We utilise the notation $\E[\ldots\vert\mathcal{F}_t]=\E_t[\ldots]$.
In the bottom-up setting, we directly model the short risk-free rate $\{r_t\}$ in the manner of the right-hand side in \qr{bond&rate}, i.e.
\begin{equation}\label{rateBU}
r_t= -\dfrac{\dot{c_1} (t)+ \dot{b_1}(t)A_t^{(1)}}{c_1(t) + b_1(t)A_t^{(1)}},
\end{equation}
by postulating (i) non-increasing deterministic functions $b_1(t)$ and $c_1(t)$ with $c_1(0)=1$ (later $c_1(t)$ will be seen to coincide with $\Bd_{0t}$), and (ii) an $(\{\Fc_t\},\M)$-martingale $\{A^{(1)}_t\}$ with $A^{(1)}_0=0$ such that 
\begin{equation}\label{h-proc}
h_t=c_1(t)+b_1(t)A^{(1)}_t
\end{equation}
is a positive $(\{\mathcal{F}_t\},\mathbb{M}$)-supermartingale for all $t>0$.
\begin{Example}\em\label{e:theele}
Let $\theC_{t}=\S^{(1)}_t -1,$ where
$\{S^{(1)}_t\}$ is a positive $\M$-martingale with $S^{(1)}_0 =1$; for example exponential L\'evy martingales. The supermartingale (\ref{h-proc}) is positive for any given $t$ if 
$0<\theb(t) \leq  \thea(t)$. 
\end{Example}
\noindent
Associated with the supermartingale (\ref{h-proc}), we characterise the (risk-neutral) pricing measure $\mathbb{Q}$ by the  $\M$-density process $\{\mu_t\}_{0\le t\le T}$, given by
\begin{equation}\label{MQ}
\mu_t=\Frac{\rd\mathbb{Q}}{\rd\mathbb{M}} 
 \Big|_{\mathcal{F}_t} =\mathcal{E}\left(\int_0^{\cdot}\Frac{\theb (t)d\theC_t}{\thea (t) + \theb (t) \theC_{t-} } \right),
\end{equation}
which is taken to be a positive $(\{\mathcal{F}_t\},\M)$-martingale.
%check
Furthermore, we denote by $D_t=\exp\left(-\int_0^t\,r_s\,ds\right)$ the discount factor associated with the risk-neutral measure $\Q$.

\bl \label{lem:p} 
$\thep= D_t\,  \mu_t$. 
\el
\proof The Ito semimartingale formula applied to $\varphi(t,\theC_{t})=\ln(\thea (t) + \theb (t)\theC_{t})=\ln(h_{t})$ and to ${\ln}(D_t\mu_t)$ gives the following relations: 
\begin{eqnarray} 
d\ln\left(\thea (t) + \theb (t)\theC_{t}\right)
&=&-r_t dt + \Frac{\theb (t) d\theC_{t}}{\thea (t) + \theb (t)\theC_{t-}}-\Frac{ \theb^2 (t) 
d[ \theC,\theC]^c_t}{2(\thea (t) + \theb (t)\theC_{t-})^2}\nonumber\\
&+&d\sum_{s\leq t}\left(\Delta\ln\big(\thea (t) + \theb (t)\theC_{t}\big)-\Frac{\theb (t) \Delta\theC_{t} }{\thea (t) + \theb (t)\theC_{t-}}\right),\nonumber\\
\end{eqnarray}
where \qr{rateBU} was used in the first line,
and 
\begin{eqnarray}
d\ln(D_t\mu_t)&=&d\ln D_t + d\,{\ln}\mu_t\nonumber\\
&=&-r_t dt + \Frac{d\mu_{t}}{\mu_{t-}}- \Frac{d[  \mu,\mu]^c_t}{2(\mu_{t-})^2}+d\sum_{s\leq t}\left(\Delta\ln(\mu_t)- \Frac{\Delta\mu_t}{\mu_{t-}}  \right)\nonumber\\
&=&
-r_t dt +  \Frac{\theb (t)d\theC_t}{\thea (t) + \theb (t) \theC_{t-} }-\Frac{ \theb^2 (t) 
d[ \theC,\theC]^c_t}{2(\thea (t) + \theb (t)\theC_{t-})^2}\nonumber\\
&&\hspace{4.25cm}+d\sum_{s\leq t}\left(\Delta\ln (\mu_t )-\Frac{\theb (t)\Delta\theC_{t}  }{\thea (t) + \theb (t)\theC_{t-}}\right)
\end{eqnarray}
where 
\bel
\Delta\ln\left(\mu_t\right)&=\ln\left( \Frac{\mu_t}{\mu_{t-}} \right)=\ln\left(1+\Frac{\theb (t) \Delta\theC_{t} }{\thea (t) + \theb (t)\theC_{t-}}\right)
=\ln\left(\Frac{\thea (t) + \theb (t)\theC_{t} }{\thea (t) + \theb (t)\theC_{t-}}\right)\\&=\Delta\ln\left(\thea (t) + \theb (t)\theC_{t}\right). 
\eel
Therefore, $d\ln(h_t)=d\ln (D_t\mu_t)$. Moreover, $h_0=D_0\mu_0=1.$ Hence
$h_t=D_t\mu_t.$ \Finproof

It then follows that the price process of the OIS discount bond with maturity $T$ can be expressed by
\beql{e:z}
&\Bd_{tT}=\EQ_ t \left[\Frac{D_T}{D_t} \right] = \dfrac{1}{D_t\,\mu_t}\E^{\M}\left[D_T\,\mu_T\,\vert\,\Fc_t\right] 
=\EM_ t \left[\Frac{\thep_T  }{\thep_t } \right] =\Frac{\thea (T) + \theb (T)\theC_{t}}{\thea (t) + \theb (t)\theC_{t}},
\eeql
for $0\leq t\leq T$, Thus, the process $\{h_t\}$ plays the role of the pricing kernel associated with the OIS market under the measure $\M$. In particular, we note that $c_1(t)=\Bd_{0t}$ for $t\in[0,T]$ and $r_t=-\left(\partial_T \ln \Bd_{tT}\right)_{|T=t}\geq  0$. A construction inspired by the above formula for 
the OIS bond leads to the rational model for the LIBOR prevailing over the interval $[T_{i-1}, T_i)$. The $\Fc_{T_{i-1}}$-measurable spot LIBOR rate $L(T_i;T_{i-1},T_i)$ is modelled in terms of $\{A^{(1)}_t\}$ and, in this paper, at most two other $\M$-martingales $\{A^{(2)}_t\}$ and $\{A^{(3)}_t\}$ evaluated at $T_{i-1}$:
\begin{equation}\label{LIBOR_model}
L(T_i;T_{i-1},T_i)=\dfrac{L(0;T_{i-1},T_i) + b_2(T_{i-1},T_i)A_{T_{i-1}}^{(2)}+ b_3(T_{i-1},T_i)A_{T_{i-1}}^{(3)}}{\Bd_{0T_i} + b_1(T_i)A_{T_{i-1}}^{(1)}}.
\end{equation}
The (forward) LIBOR process is then defined by an application of the risk-neutral valuation formula (which is equivalent to the pricing formula (\ref{pf-formula}) under $\PR$) as follows. For $t\leq T_{i-1}$ we let
\begin{eqnarray}
L(t;T_{i-1},T_i)&=&\dfrac{1}{D_t}\E_t^{\Q}\left[D_{T_i}\,L(T_i;T_{i-1},T_i)\right]
=\E_t^{\M}\left[\frac{D_{T_i}\,\mu_{T_i}}{D_t\,\mu_t}\,L(T_i;T_{i-1},T_i)\right]\\
&=&\E_t^{\M}\left[\dfrac{\E_{T_{i-1}}^{\M}[h_{T_i}]L_{T_i;T_{i-1},T_i}}{h_t}\right], 
 \end{eqnarray}
and thus, by applying (\ref{h-proc}) and (\ref{LIBOR_model}),
\begin{equation}\label{LIBOR}
L(t;T_{i-1},T_i)=\dfrac{L(0;T_{i-1},T_i) + b_2(T_{i-1},T_i)A_t^{(2)} + b_3(T_{i-1},T_i)A_t^{(3)}}{\Bd_{0t} + b_1(t)A_t^{(1)}}.
\end{equation}
Hence, we recover the same model (and expression) as in (\ref{LIBOR-process}). The LIBOR models (\ref{LIBOR}) (or (\ref{LIBOR-process})) are compatible with an HJM multi-curve setup where, in the spirit of \cite{hjm}, the initial term structures $\Bd_{0 T_i}$ and $L(0;T_{i-1},T_i)$ are fitted by construction.
 
\begin{Example}\em\label{e:theelecont} 
Let $A^{(i)}_{t}=\S^{(i)}_t -1,$ where
$\S^{(i)}_t$ is a positive $\M$-martingale with $\S^{(i)}_0 =1$. For example, one could consider a unit-initialised exponential L\'evy martingale defined in terms of a function of an $\M$-L\'evy process $\{X^{(i)}_t\}$, for $i=2,3$. Such a construction produces non-negative LIBOR rates if 
\beql{e:nonne}
0\leq b_2(T_{i-1},T_i)  +  b_3(T_{i-1},T_i) \leq  L(0;T_{i-1},T_i).
\eeql
If this condition is not satisfied, then the LIBOR model may be viewed as a shifted model, in which the LIBOR rates may become negative with positive probability. For different kinds of shifts used in the multi-curve term structure literature we refer to, e.g., \citeN{Mercurio2010a} or \citeN{MoreniPallavicini10}.
\end{Example}
%%%%%%%%%%%%%%%%%%%%%%%%%%%%%%%%%%%%%%%%%%%%%%%%
\section{Clean valuation}\label{s:pg}

The next questions we address are centred around the pricing of LIBOR derivatives and their calibration to market data, especially LIBOR swaptions, which are the most liquidly traded (nonlinear) interest rate derivatives. Since market data typically reflect prices of fully collaterallised transactions, which are funded at a remuneration rate of the collateral that is best proxied by the OIS rate, we consider in this section, in the perspective of model calibration, 
clean valuation ignoring counterparty risk and assume funding at the rate $r_t.$

An interest rate swap (see, e.g., \cite{bm}) is an agreement between two counterparties, where one stream of future interest payments is exchanged for another based on a specified nominal amount $N$. A popular interest rate swap is the exchange of a fixed rate (contractual swap spread) against the LIBOR at the end of successive time intervals $[T_{i-1},T_i]$ of length $\delta$. Such a swap can also be viewed as a collection of $n$ forward rate agreements. The swap price $Sw_t$ at time $t\leq T_0$ is given by the following model-independent formula:
\begin{equation*}
Sw_t = N\delta\sum_{i=1}^n   [ L(t;T_{i-1},T_i)-K\Bd_{tT_i} ].
\end{equation*}

A swaption is an option between two parties to enter a swap at the  expiry date $T_k$ (the maturity date of the option). Its price at time $t\leq T_k$ is given by the following $\M$-pricing formula:
\begin{align}\label{e:summa}
Swn_{tT_k} &=\dfrac{\Nd  }{h_t}\E^{\M}[h_{T_k} (Sw_{T_k})^+|\Fc_t]\nonumber\\
&=\dfrac{\Nd }{h_t}\E^{\M}\left[h_{T_k} \left(\sum_{i=k+1}^n   [ L(T_k;T_{i-1},T_i) -K\Bd_{T_kT_i}]\right)^+\Big|\Fc_t\right] \nonumber\\
&=\Frac{\Nd  }{\Bd_{0t}+b_1 A_{t}^{(1)}} \E^{\M}\Big[\Big(\sum_{i=k+1}^{m}\big[L(0;T_{i-1},T_i)+b_2 (T_{i-1},T_i)A_{T_k}^{(2)}+b_3 (T_{i-1},T_i)A_{T_k}^{(3)}\nonumber\\
&\hspace{5cm}-K(\Bd_{0T_i}+b_1(T_i)A_{T_k}^{(1)})\big]\Big)^{+} \Big|\Fc_t\Big]
\end{align} 
using the formulae \eqref{e:z} and \eqref{LIBOR} for $\Bd_{T_kT_i}$ and $L(T_k;T_{i-1},T_i)$. In particular, the swaption prices at time $t=0$  can be rewritten by use of $A^{(i)}_{t}=\S^{(i)}_t -1$ so that
\beql{e:s}
Swn_{0T_k} 
&= {\Nd } \,
\E^{\M}\left[\left(c_2 A_{T_k}^{(2)}+c_3 A_{T_k}^{(3)}-c_1 A_{T_k}^{(1)} +c_0\right)^{+}\right]\\
&=
{\Nd } \,
\E^{\M}\left[\left(c_2 S_{T_k}^{(2)}+c_3 S_{T_k}^{(3)}  -c_1 S_{T_k}^{(1)}+\tilde{c}_0 \right)^{+}\right],
\eeql
where
\bel
&c_2=\sum_{i=k+1}^{m}  b_2 (T_{i-1},T_i),\quad  c_3=\sum_{i=k+1}^{m}  b_3 (T_{i-1},T_i), \quad c_1=K\sum_{i=k+1}^{m}  b_1(T_i),&\\
&c_0=\sum_{i=k+1}^{m}  [L(0;T_{i-1},T_i)-K \Bd_{0T_i}],\quad \tilde{c}_0=c_0+c_1- c_2 -c_3.& 
\eel
As we will see in several instance of interest, these expectations can be computed efficiently with high accuracy by various numerical schemes.
\begin{Remark}\em
The advantages of modelling the LIBOR process $\{L(t;T_{i-1},T_i)\}$ by a rational function of which denominator is the discount factor (pricing kernel) associated with the employed pricing measure (in this case $\M$) are: (i) The rational form of $\{L(t;T_{i-1},T_i)\}$ and also of $\{\Bd_{tT_i}\}$ produces, when multiplied with the discount factor $\{h_t\}$, a linear expression in the $\M$-martingale drivers $\{A^{(i)}_t\}$. This is in contrast to other akin pricing formulae in which the factors appear as sums of exponentials, see e.g. \cite{CrepeyGrbacNgorSkovmand13}, Equation (33). (ii) The dependence structure between the LIBOR process and the OIS discount factor $\{h_t\}$---or the pricing kernel $\{\pi_t\}$ under the $\PR$-measure---is clear-cut. The numerator of $\{L(t;T_{i-1},T_i)\}$ is driven only by idiosyncratic stochastic factors that influence the dynamics of the LIBOR process. We may call such drivers the ``LIBOR risk factors''. Dependence on the ``OIS risk factors'', in our model example $\{A^{(1)}_t\}$, is produced solely by the denominator of the LIBOR process. (iii) Usually, the FRA process $K_t =L(t;T_{i-1},T_i)/\Bd_{t T_i}$ is modelled directly and more commonly applied to develop multi-curve frameworks. With such models, however, it is not guaranteed that simple pricing formulae like (\ref{e:summa}) can be derived. We think that the ``codebook'' (\ref{LIBOR-proc}), and (\ref{LIBOR}) in the considered example, is more suitable for the development of consistent, flexible and tractable multi-curve models.      
\end{Remark}
%%%%%%%%%%%%%%%%%%%%%%%%%%%%%%%%
\subsection{Univariate Fourier pricing}
Since in current markets there are no liquidly-traded OIS derivatives and hence no useful data is available, a pragmatic simplification is to assume deterministic OIS rates $r_t$. That is to say $A_t^{(1)}=0$, and hence $b_1(t)$ plays no role either, so that it can be assumed equal to zero. Furthermore, for a start, we assume $A_t^{(3)}=0$ and $b_3(t)=0$, and \eqref{e:s} simplifies to
$$
Swn_{0 T_k} 
=  {\Nd}\,\E^{\M}\left[\left(c_2 A_{T_k}^{(2)}+c_0\right)^{+}\right]=
\Nd\,\E^{\M}\left[\left(c_2 S_{T_k}^{(2)}  { +} \tilde{c}_0\right)^{+}\right],
$$
where here $\tilde{c}_0=c_0  \s{ -}c_2$. For $\tilde{c}_0>0$ the price is simply  $ Swn_{0 T_k} =  {\Nd}c_0$. For $\tilde{c}_0<0$, and in the case of an exponential-L\'evy martingale model with
$$ 
\S^{(2)}_t=\e^{\theX ^{(2)}_t-t\,\psi_2(1)},
$$
where $\{X_t^{(2)}\}$ is a L\'evy process with cumulant $\psi_2$
 such that
\begin{equation}\label{e:cum}
\E\left[\e^{z \theX ^{(2)}_t}\right] = \exp \left[t \psi_2(z) \right],
\end{equation}
we have
\begin{equation}\label{EGP}
Swn_{0T_k} = \Frac{\Nd}{2\pi} 
\int_\R  \Frac{{\tilde{c}_0}^{\ 1-iv-R}\ M_{T_k}^{(2)}(R+iv)}{(R+iv)(R+iv-1)}dv , 
\end{equation}
where
$$M_{T_k}^{(2)}(z)=\e^{T_k \psi_2(z)+z\big(\ln(c_2) - \psi_2 (1)\big)} $$  and $R$ is an arbitrary constant ensuring finiteness of $ M_{T_k}^{(2)}(R+iv)$ for $v\in\R$. For details concerning (\ref{EGP}), we refer to, e.g., \citeN{EberleinGlauPapapantoleon}.
%%%%%%%%%%%%%%%%%%%%%%%%%%%%%%%%%%%%%%%%%%%%%%%%
\subsection{One-factor lognormal model}\label{ss:ln1}
In the event that \s{$\{A^{(1)}_t\}=\{A^{(3)}_t\}=0$} and $\{A_t^{(2)}\}$ is of the form
\begin{equation}\label{lognormal2}
A^{(2)}_t = \exp \left(a_2 X^{(2)}_t - \Frac{1}{2}a_2^2t\right)-1,  
\end{equation}
where $\{X^{(2)}_t\}$ is a standard Brownian motion and $a_2$ is a real constant,
it follows from simple calculations that the swaption price is given, for $\tilde{c}_0=c_0-c_2,$
by
\begin{align}\label{e:swnln1}
Swn_{0 T_k} 
=&  {\Nd}\,\E^{\M}\left[\left(c_2 A_{T_k}^{(2)} 
{ +}c_0\right)^{+}\right]\\
=&\Nd\left(c_2 \Phi\left(\frac{\frac 12a_2^2T-\ln(\tilde{c}_0/c_2)}{a_2\sqrt{T}}\right)+\tilde{c}_0\Phi\left(\frac{-\frac 12a_2^2T-\ln(\tilde{c}_0/c_2)}{a_2\sqrt{T}}\right)\right),
\end{align}
where $\Phi(x)$ is the standard normal distribution function.

\subsection{Two-factor lognormal model}\label{ss:ln2}
We return to the price formula (\ref{e:s}) and consider the case where the martingales $\{A_t^{(i)}\}$ are given, for $i=1,2,3,$ by
\begin{align}\label{lognormal23}
&A^{(i)}_t = \exp \left(a_i X^{(i)}_t - \Frac{1}{2}a_i^2t\right)-1 ,&    
\end{align}
for real constants $a_i$
and standard Brownian motions $\{X^{(1)}_t\}=\{X^{(3)}_t\}$ and $\{X^{(2)}_t\}$  with correlation $\rho$. Then it follows that
\begin{equation}\label{e:s2}
Swn_{0T_k} 
=\E^{\M}\left[\left( c_2\e^{X\sqrt{T_k}a_2 -\frac12 a_2^2T_k}  +c_3\e^{Y\sqrt{T_k}a_3 -\frac12 a_3^2T_k}  -c_1\e^{Y\sqrt{T_k}a_1 -\frac12 a_1^2T_k}+\tilde{c}_0 \right)^{+}\right],
\end{equation}
where $X\sim\mathcal{N}(0,1)$, $Y\sim\mathcal{N}(0,1)$, $(X|Y)=y \sim\mathcal{N}(\rho y,(1-\rho^2))$. Hence,
\begin{align*}
Swn_{0T_k}&=\int_{-\infty}^{\infty}\int_{-\infty}^{\infty}(c_2\e^{x\sqrt{T_k}a_2 -\frac12 a_2^2T_k} -K(y))^+f(x|y)f(y)dxdy\\
&=\int_{K(y)>0}\left(\int_{-\infty}^{\infty}(c_2\e^{x\sqrt{T_k}a_2 -\frac12 a_2^2T_k} -K(y))^+f(x|y)dx\right)f(y)dy\\
&\hspace{1.5cm}+\int_{K(y)<0}\left(\int_{-\infty}^{\infty}(c_2\e^{x\sqrt{T_k}a_2 -\frac12 a_2^2T_k} -K(y))^+f(x|y)dx\right)f(y)dy,
\end{align*}
where 
\bel
&K(y) = c_1(\e^{a_1\sqrt{T_k}y -\frac 12 a_1^2 T_k}-1) -c_3(\e^{a_3\sqrt{T_k}y -\frac 12 a_3^2 T_k}-1) -c_0,\\
&f(y)=\frac{1}{\sqrt{2\pi}}\e^{-\frac{y^2}{2}},\\
&f(x|y)= \frac{1}{\sqrt{2\pi(1-\rho^2)}}\e^{\frac{-(x-\rho y)^2}{2(1-\rho^2)}}.
\eel
This expression can be simplified further to obtain
\begin{eqnarray*}
&&Swn_{0T_k}\\
&&=\int_{K(y)>0}\Bigg[c_2\e^{a_2\sqrt{T_k}\,\rho y +\frac12 a_2^2T_k(1-\rho^2)}\Phi\left(\frac{ \rho y +a_2\sqrt{T_k}(1-\rho^2) +\ln(c_2)-\frac12 a_2^2T_k -K(y)}{\sqrt{1-\rho^2}}\right)\\
 &&\hspace{2.25cm}-K(y)\Phi\left(\frac{ \rho y +\ln(c_2)-\frac12 a_2^2T_k -K(y)}{\sqrt{1-\rho^2}}\right) \Bigg] f(y)dy \\
 &&+\int_{K(y)<0}\left(c_2 \e^{a_2\sqrt{T_k}\rho(y-\frac 12 a_2\sqrt{T_k}\rho}-K(y)\right) f(y)dy.
\end{eqnarray*}
The calculation of the swaption price is then reduced to calculating two one-dimensional integrals. Since the regions of integration  are not explicitly known, one has to numerically solve for the roots of $K(y)$, which may have up to two roots. Nevertheless a full swaption smile can be calculated in a small fraction of a second by means of this formula.
\section{Calibration}\label{s:cal}
The counterparty-risk valuation adjustments, abbreviated by XVAs (CVA, DVA, LVA, etc.), can be viewed as long-term options on the underlying contracts. For their computation, the effects by the volatility smile and term structure matter. Furthermore, for the planned XVA computations of the multi-curve products in Section~\ref{s:crisk}, it is necessary to calibrate the proposed pricing model to financial instruments with underlying tenors of $\delta=3$m and $\delta=6$m (the most liquid tenors). Similar to \cite{CrepeyGrbacNgorSkovmand13}, we make use of the following EUR market Bloomberg data of January 4, 2011 to calibrate our model: EONIA, three-month EURIBOR and six-month EURIBOR initial term structures on the one hand, and three-month and six-month tenor swaptions on the other. As in the HJM framework of \cite{CrepeyGrbacNgorSkovmand13}, to which the reader is referred for more detail in this regard, the initial term structures are fitted by construction in our setup. Regarding swaption calibration, at first, we calibrate the non-maturity/tenor-dependent parameters to the swaption smile for the 9$\times$1 years swaption with a three-month tenor underlying.  
The market smile corresponds to a vector of strikes $[-200 ,-100, -50, -25, \linebreak 0 ,25, 50, 100, 200]$ bps around the underlying swap spread. Then, we make use of at-the-money swaptions on three and six-month tenor swaps all terminating at exactly ten years, but with maturities from one to nine years. This co-terminal procedure is chosen with a view towards the XVA application in  Section~\ref{s:crisk}, where a basis swap with a ten-year terminal date is considered. 

In particular, in a single factor $\{A^{(2)}_t\}$ setting:
\begin{enumerate}
\item First, we calibrate the parameters of the driving martingale $\{A^{(2)}_t\}$ to the smile of the 9$\times$1 years swaption with tenor $\delta=3$m. This part of the calibration procedure gives us also the values of $b_2(9,9.25)$, $b_2(9.25,9.5)$, $b_2(9.5,9.75)$ and $b_2(9.75,10),$ which we assume to be equal.
 \item Next, we consider the co-terminal, $\Delta\times(10- \Delta)$, ATM swaptions with $ \Delta=$ 1, 2,$\dots$, 9 years. These are available written on the three and six-month rates.
We calibrate the remaining values of $b_2$ one maturity at a time, going backwards and starting with the 8$\times$2 years for the three-month tenor and with the  9$\times$1 years for the six-month tenor. This is done assuming that the parameters are piecewise constant such that $b_2(T,T+0.25)=b_2(T+0.25,T+0.5)=b_2(T+0.5,T+0.75)=b_2(T+0.75,T+1)$ for each $T=0,1,\dots,8$ and that $b_2(T,T+0.5)=b_3(T+0.5,T+1)$ hold for each $T=0,1,\dots,9$.
\end{enumerate}
\subsection{Calibration of the one-factor lognormal model}
\begin{figure}[H] 
  \centering
 \includegraphics[width=.49\textwidth,,height=6cm]{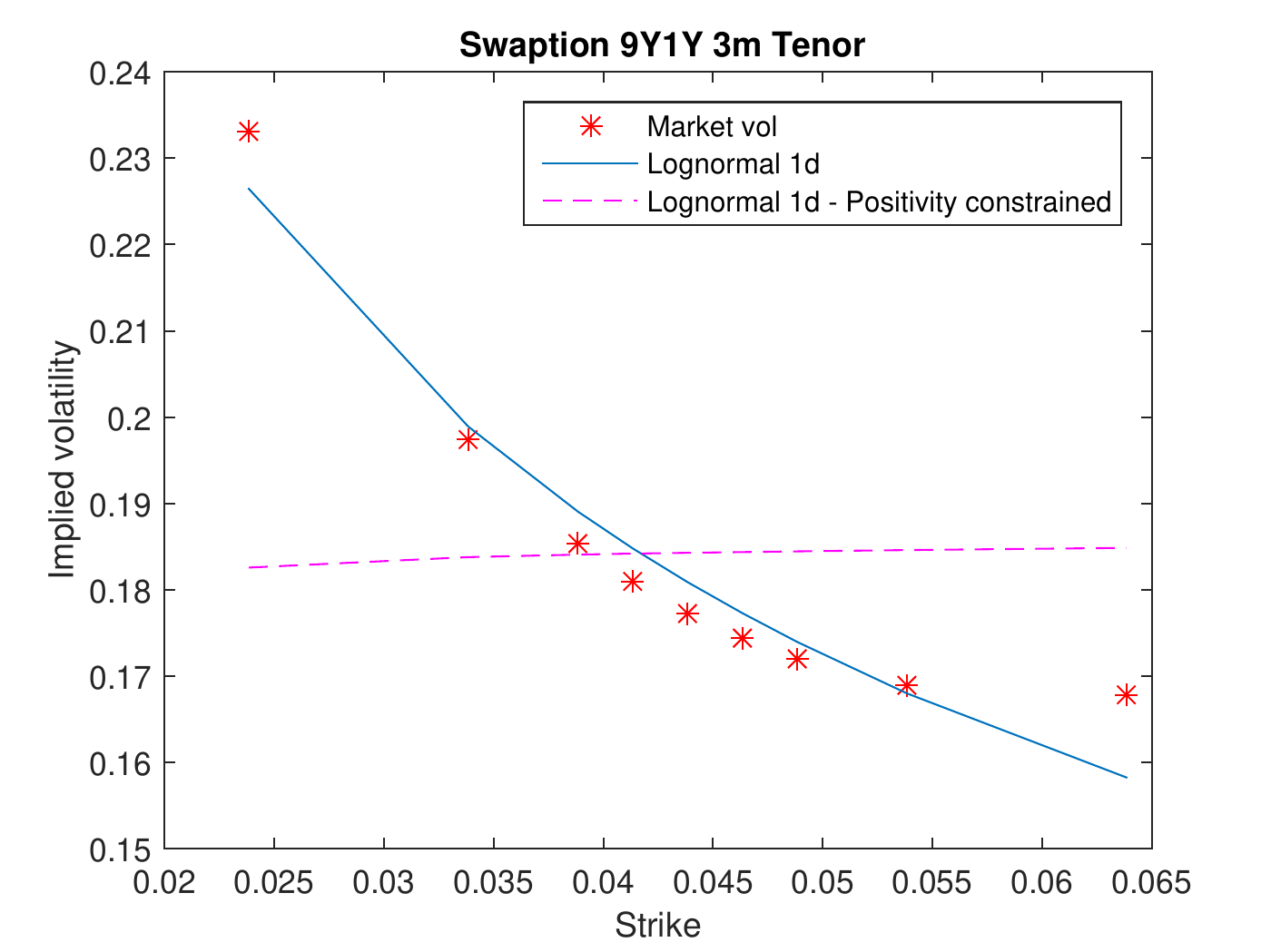}
\caption{Lognormal one-factor calibration}\label{f:Gauss1dsmile}
\end{figure}

\begin{figure}[H]
  \centering
        \includegraphics[width=.49\textwidth,,height=6cm]{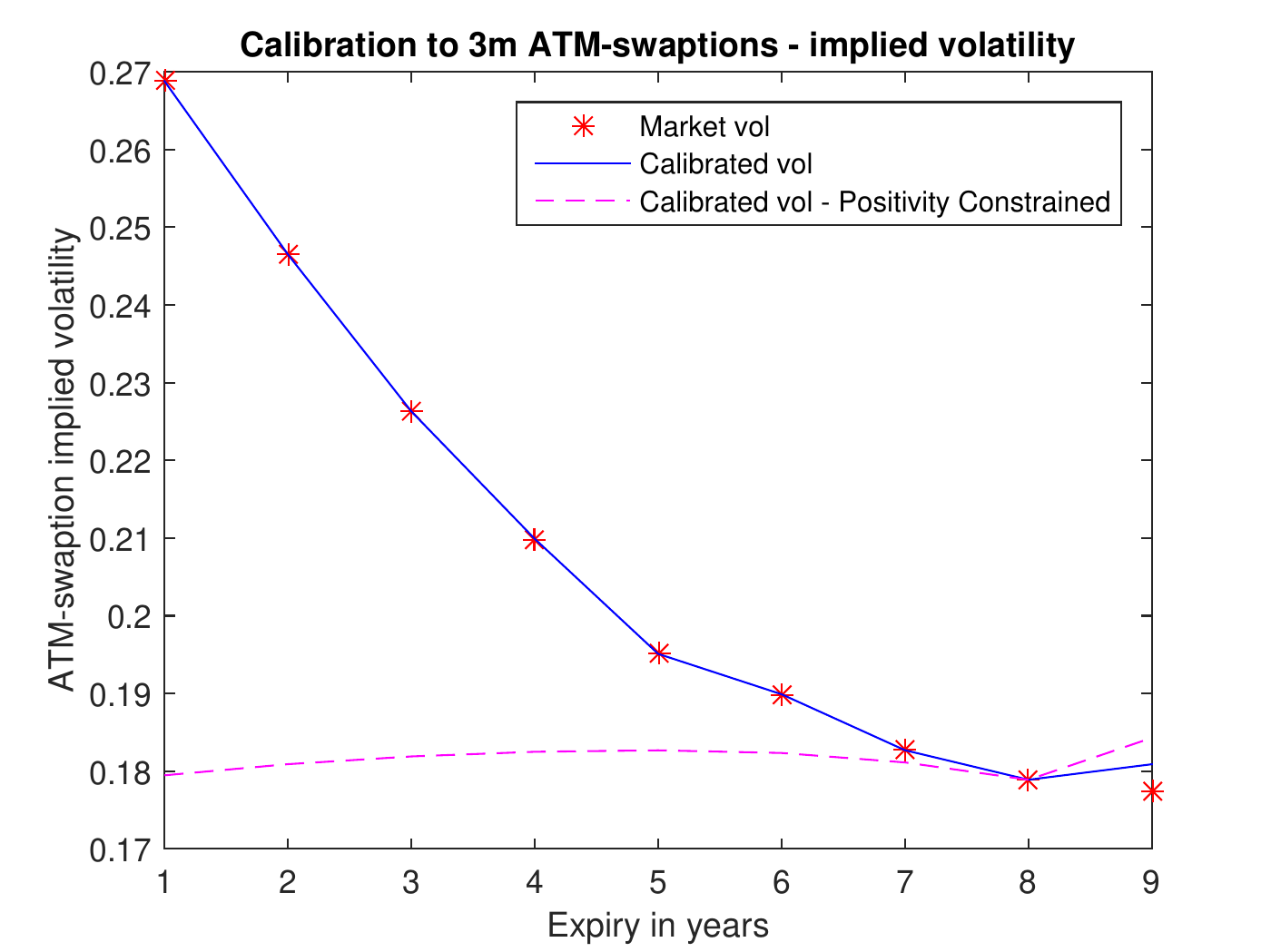}
        \includegraphics[width=.49\textwidth,,height=6cm]{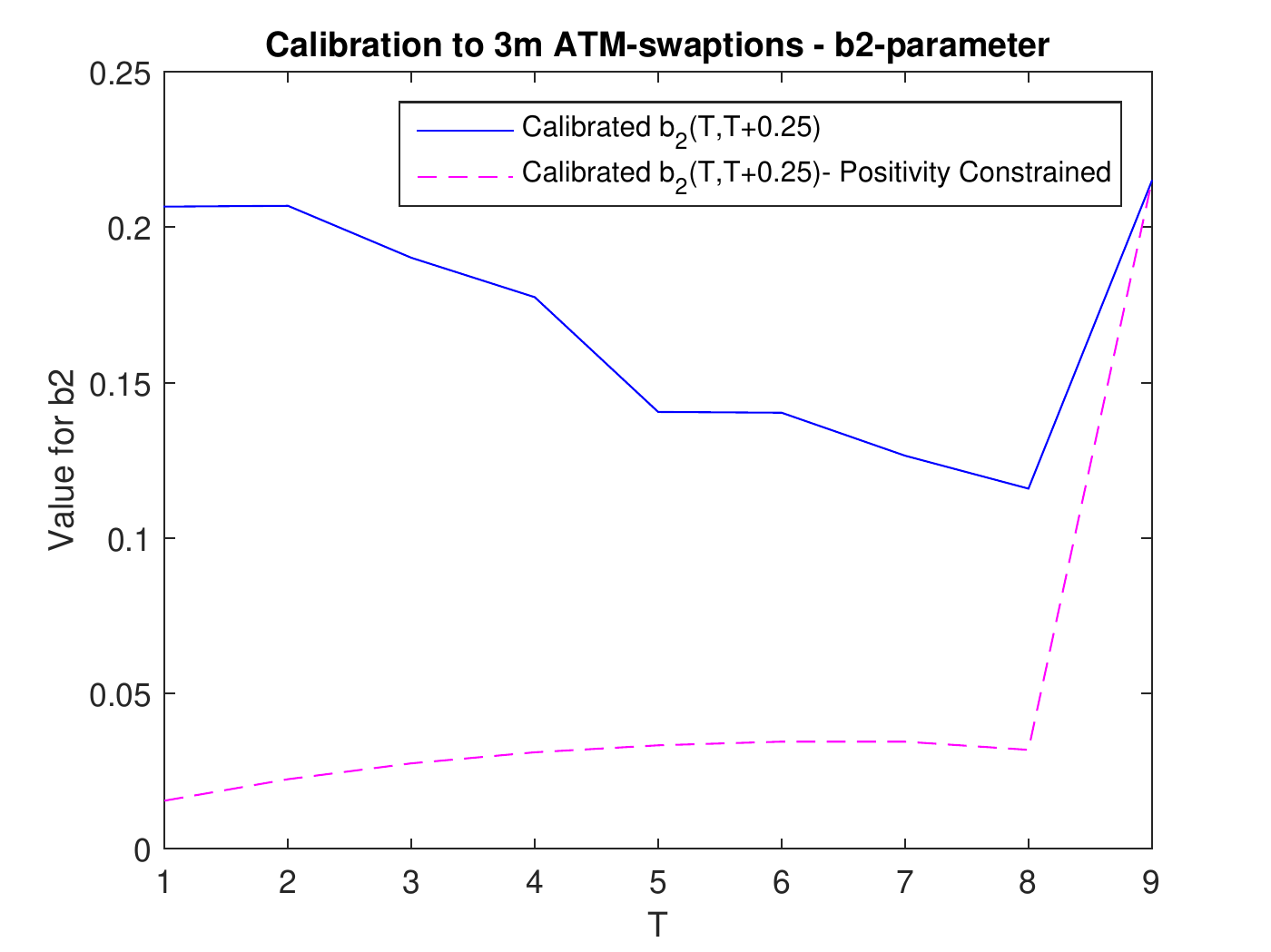}
        \includegraphics[width=.49\textwidth,,height=6cm]{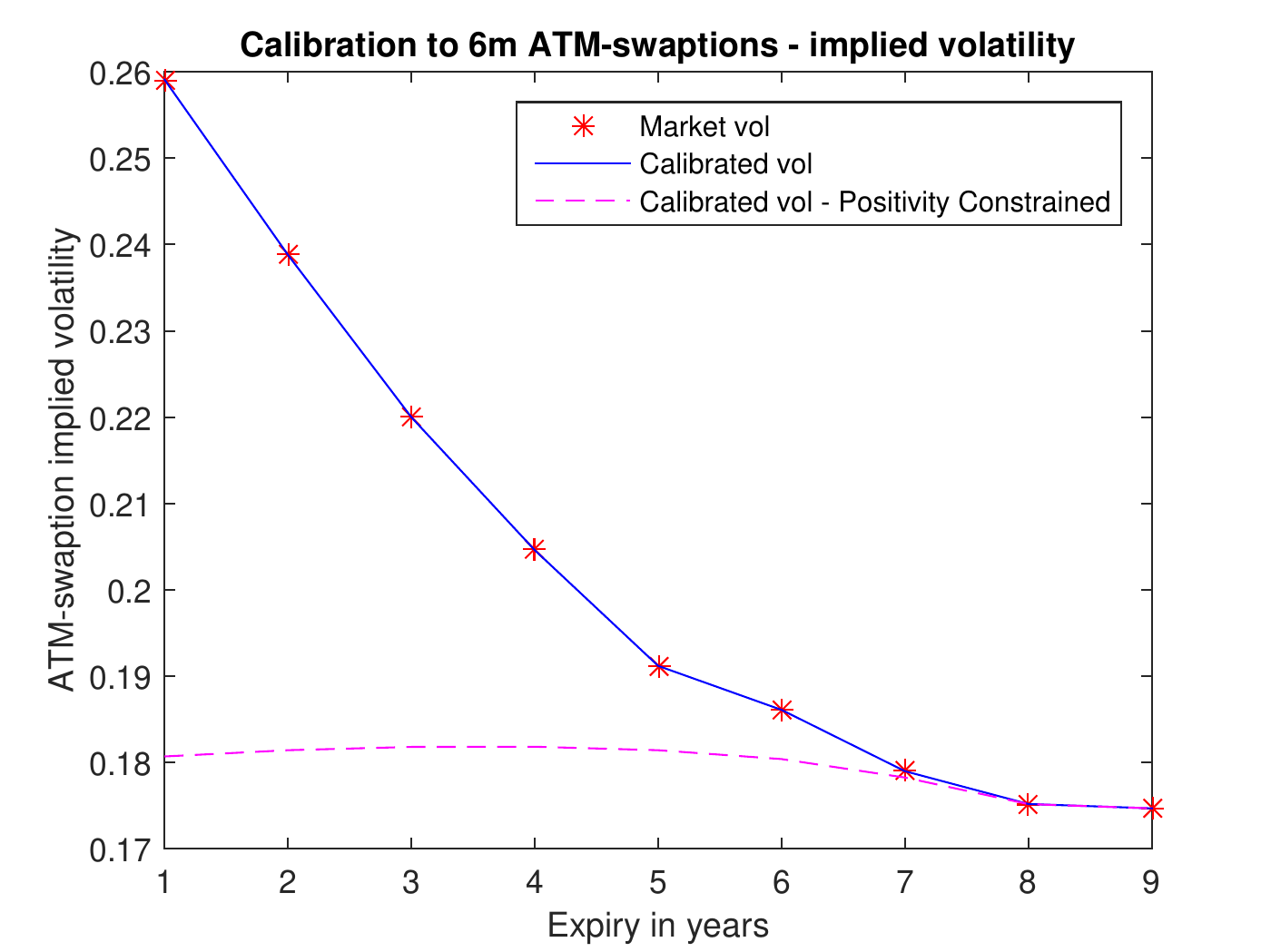}
        \includegraphics[width=.49\textwidth,,height=6cm]{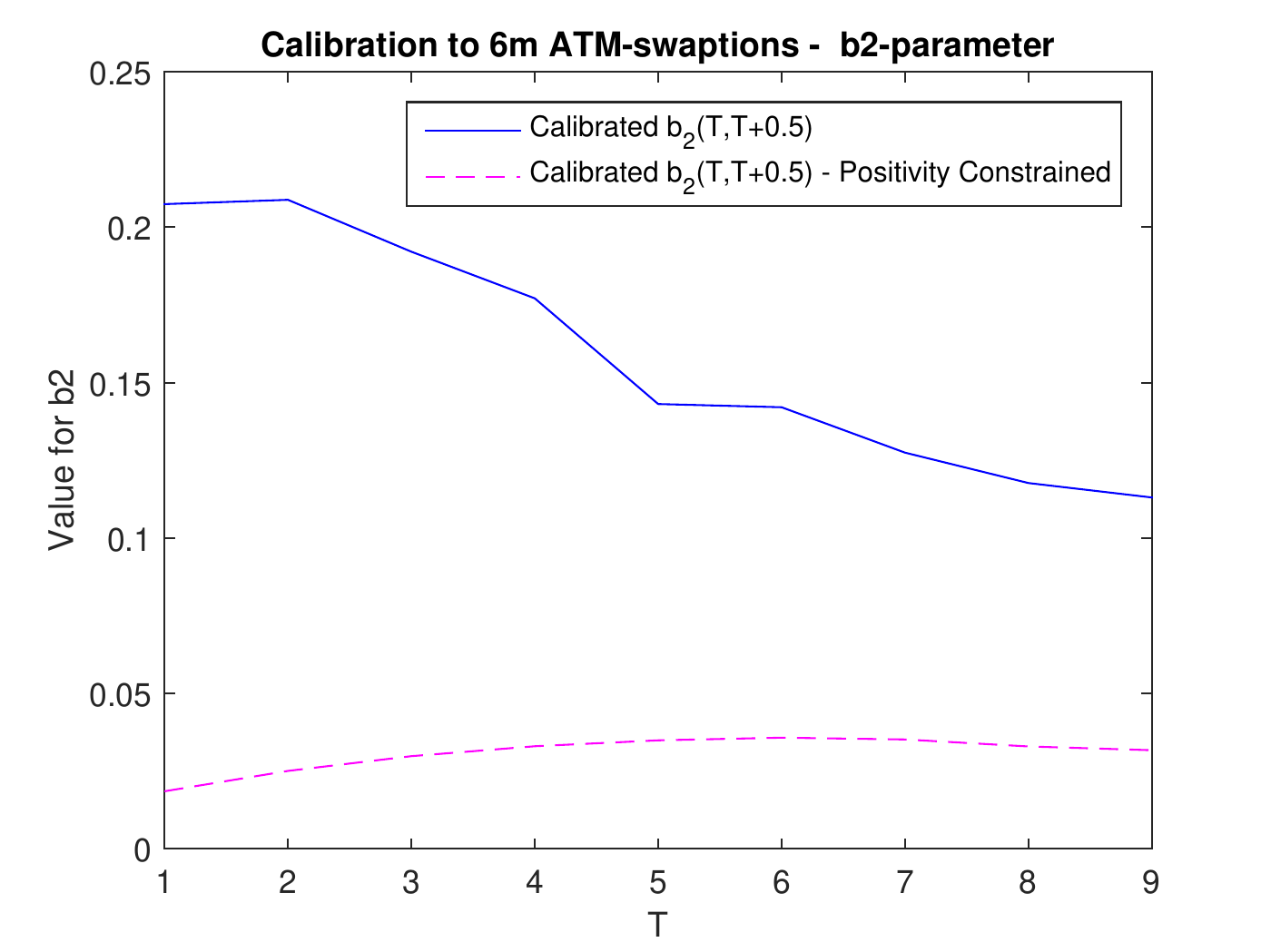}
\caption{One-Factor Lognormal calibration. {\it (Left)} Fit to ATM swaption implied volatility term structures. {\it (Right)} Calibrated values of the $b_2$ parameters. {\it (Top)} $\delta=3m$. {\it (Bottom)} $\delta=6m$.}\label{f:Gauss1dvolsandparms}
\end{figure}

In the one-factor lognormal specification of \sr{ss:ln1}, we calibrate the parameter $a_2$ and $b = b_2(9,9.25)=b_2(9.25,9.5)=b_2(9.5,9.75) =b_2(9.75,10)$ with Matlab utilising the procedure ``lsqnonlin'' based on the pricing formula \qr{e:swnln1}
(if $\tilde{c}_0<0$, otherwise $ Swn_{0 T_k} =  {\Nd}c_0$). This calibration yields:
\begin{equation*}
a_2 =    0.0537 ,b =0.1107.
\end{equation*}
Forcing positivity of the underlying LIBOR rates means, in this particular case, restricting $b\leq L(0;9.75,10)= 0.0328$ (cf. \qr{e:nonne}). The  constrained calibration yields:
\begin{equation*}
a_2 =     0.1864 ,b = 0.0328.
\end{equation*}
The two resulting smiles can be found in Figure \ref{f:Gauss1dsmile}, where we can see that the unconstrained model achieves a reasonably good calibration. However, enforcing positivity is highly restrictive since the Gaussian model, in this setting, cannot produce a downward sloping smile.
Next we calibrate the $b_2$ parameters to the ATM swaption term structures of 3 months and 6 months tenors. The results are shown in Figure \ref{f:Gauss1dvolsandparms}. When positivity is not enforced the model can be calibrated with no error to the market quotes of the ATM co-terminal swaptions. However, one can see from the figure that the positivity constraint does not allow the $b_2$ function to take the necessary values, and thus a very poor fit to the data is obtained, in particular for shorter maturities. 

\begin{figure}[H]
  \centering
        \includegraphics[width=.49\textwidth,,height=6cm]{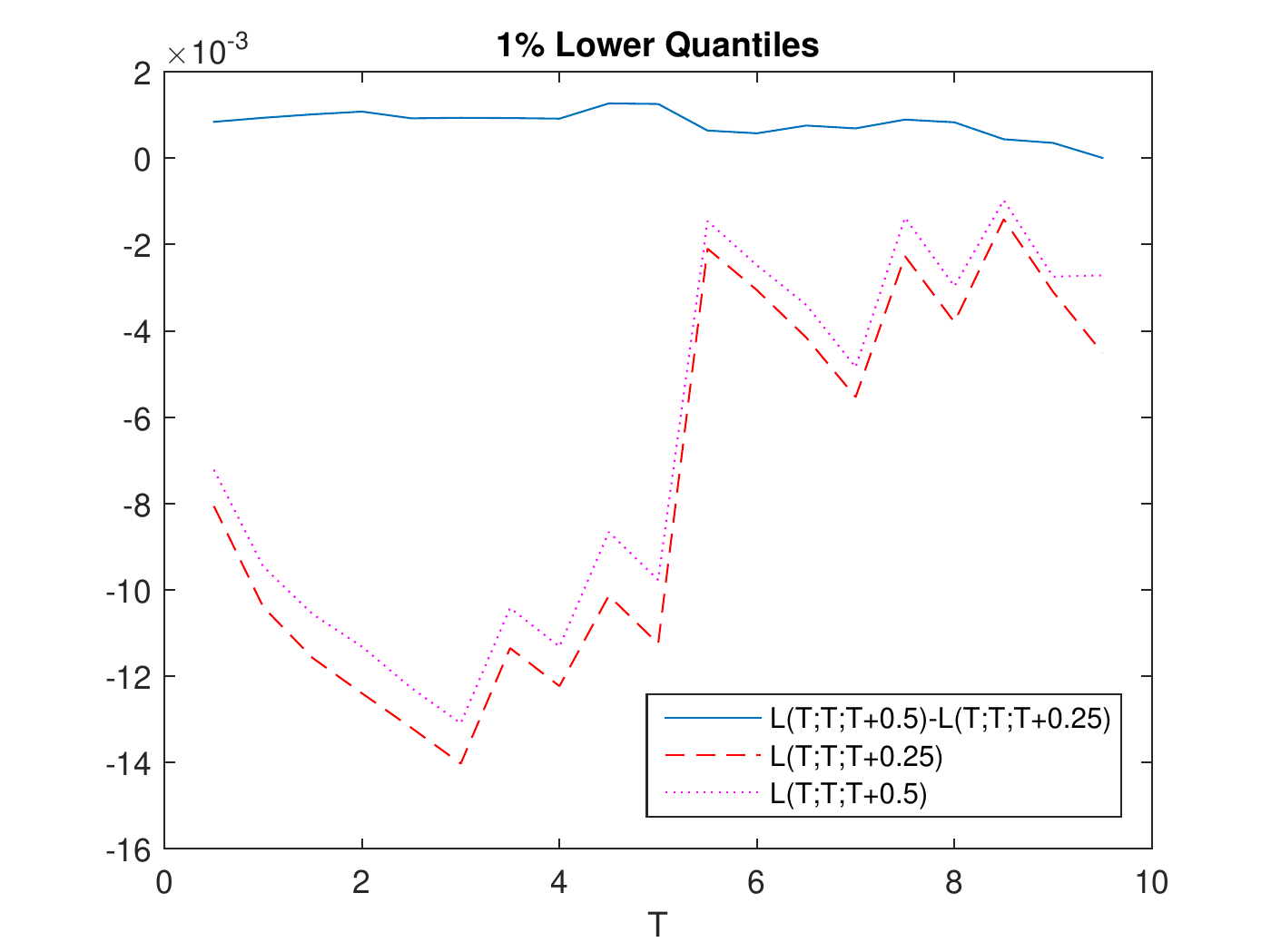}
\caption{One-Factor Lognormal calibration. 1\% lower quantiles}\label{f:Gauss1dQuantiles}
\end{figure}

With this in mind the natural question is whether the positivity constraint is too restrictive. Informal discussions with market participants reveal that positive probability for negative rates is not such a critical issue for a model. As long as the probability mass for negative values is not substantial, it is a feature that can be lived with. Indeed assigning a small probability to this event may even be realistic.\footnote{As with EONIA since the end of 2014 or Swiss rates in the crisis.}  
In order to investigate the significance of the negative rates and spreads mentioned in remark \ref{spread-pos}, we calculate lower quantiles for spot rates as well as the spot spread for the model calibrated without the positivity constraint. As Figure \ref{f:Gauss1dQuantiles} shows, the lower quantiles for the rates are of no concern. Indeed it can hardly be considered pathological that rates will be below -14 basis points with 1\% probability on a three year time horizon. Similarly, with regard to the spot spread, the lower quantile is in fact positive for all time horizons. Further calculations reveal that the probabilities of the eight year spot spread being negative is $1.1 \times 10^{-5}$ and the nine year is 0.008 -- which again can hardly be deemed pathologically high. 

We find that the model performs surprisingly well despite the parsimony of a one-factor lognormal setup. While positivity of rates and spreads are not achieved, the model assigns only small probabilities to the negatives. However, the ability of fitting the smile with such a parsimonious model is not satisfactory (cf. Fig. \ref{f:Gauss1dsmile}), which is our motivation for the next specification.
%%%%%%%%%%%%%%%%%%%%%%%%%%%%%%%%%%%%%%%%%%%%%%%%%%
%%%%%%%%%%%%%%%%%%%%%%%%%%%%%%%%%%%%%%%%%%%%%%%%%%
\subsection{Calibration of exponential normal inverse Gaussian model}\label{s:enigcal}
\begin{figure}[H] 
  \centering
 \includegraphics[width=.49\textwidth,,height=6cm]{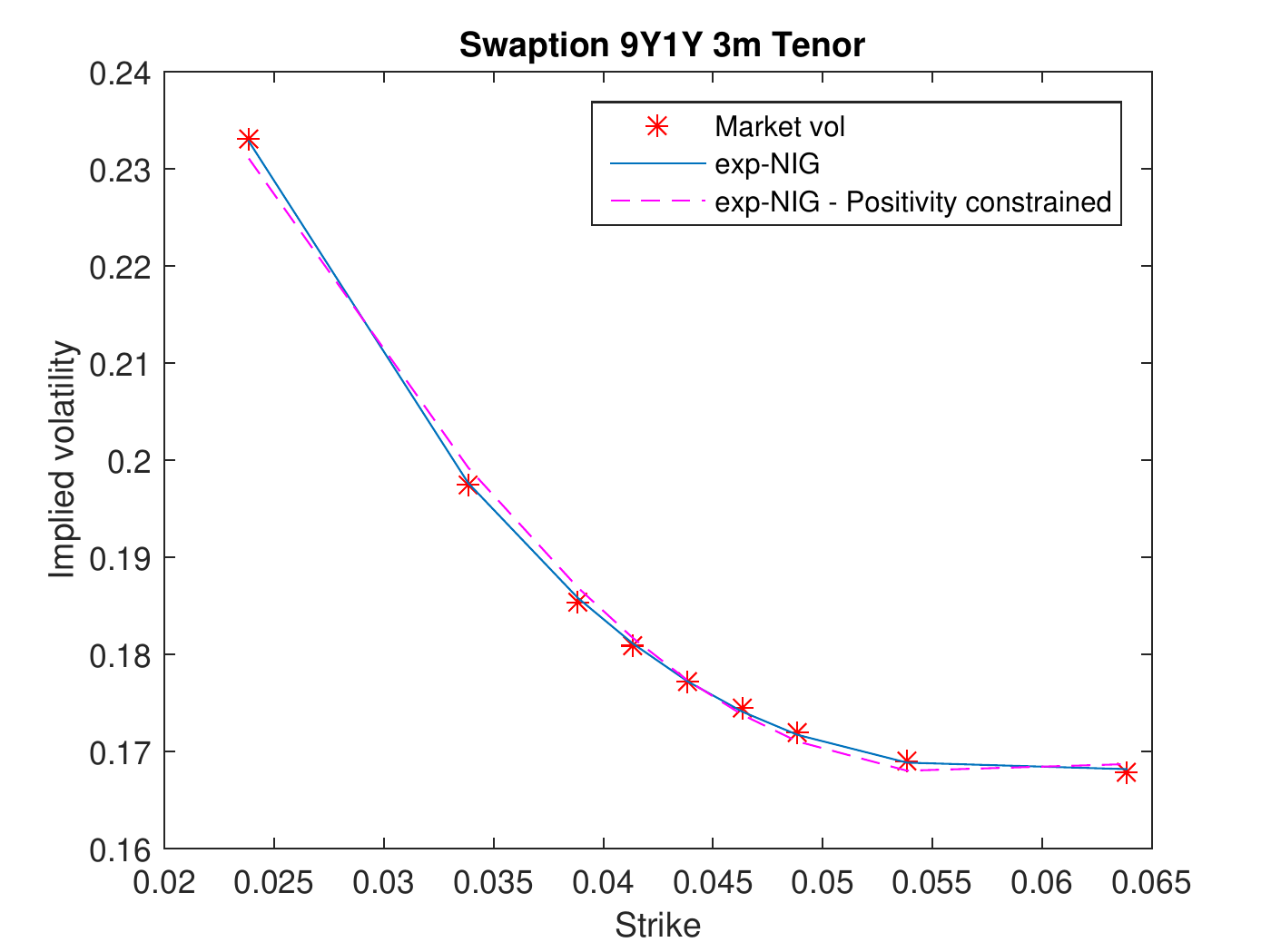}
\caption{Exponential-NIG calibration}\label{f:NIG1dSmile}
\end{figure}

\begin{figure}[htbp]
  \centering
        \includegraphics[width=.49\textwidth,,height=6cm]{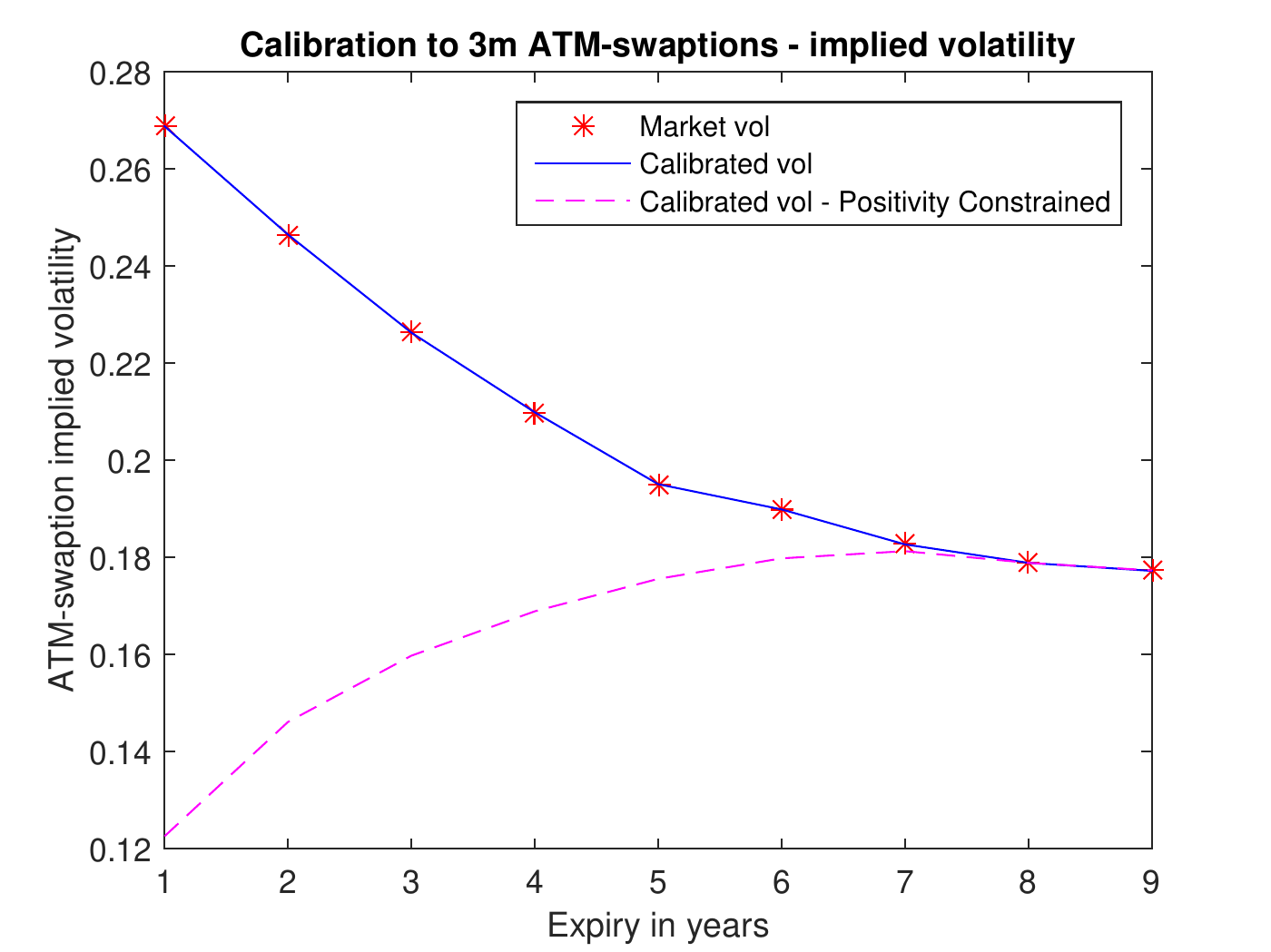}
        \includegraphics[width=.49\textwidth,,height=6cm]{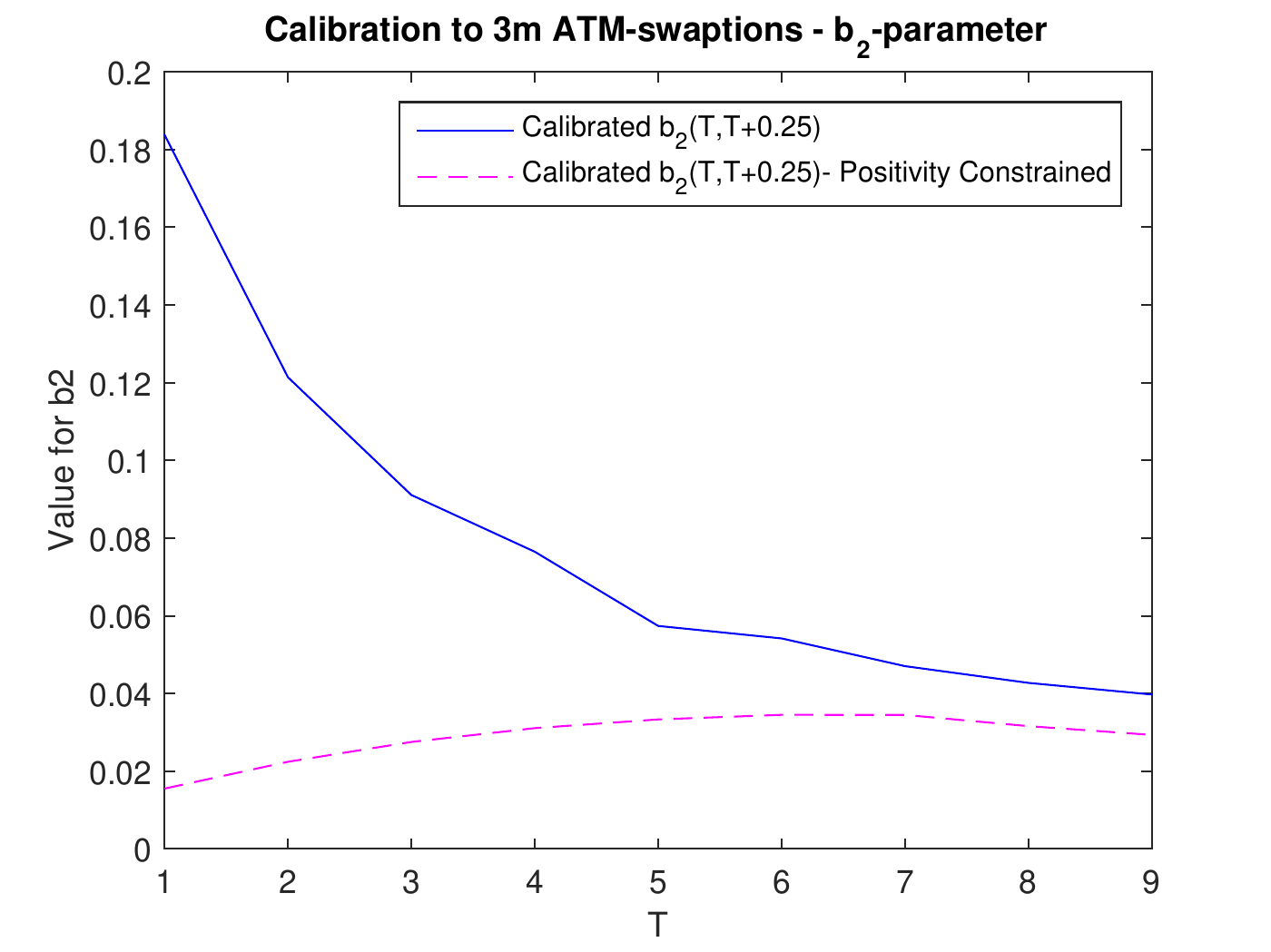}
        \includegraphics[width=.49\textwidth,,height=6cm]{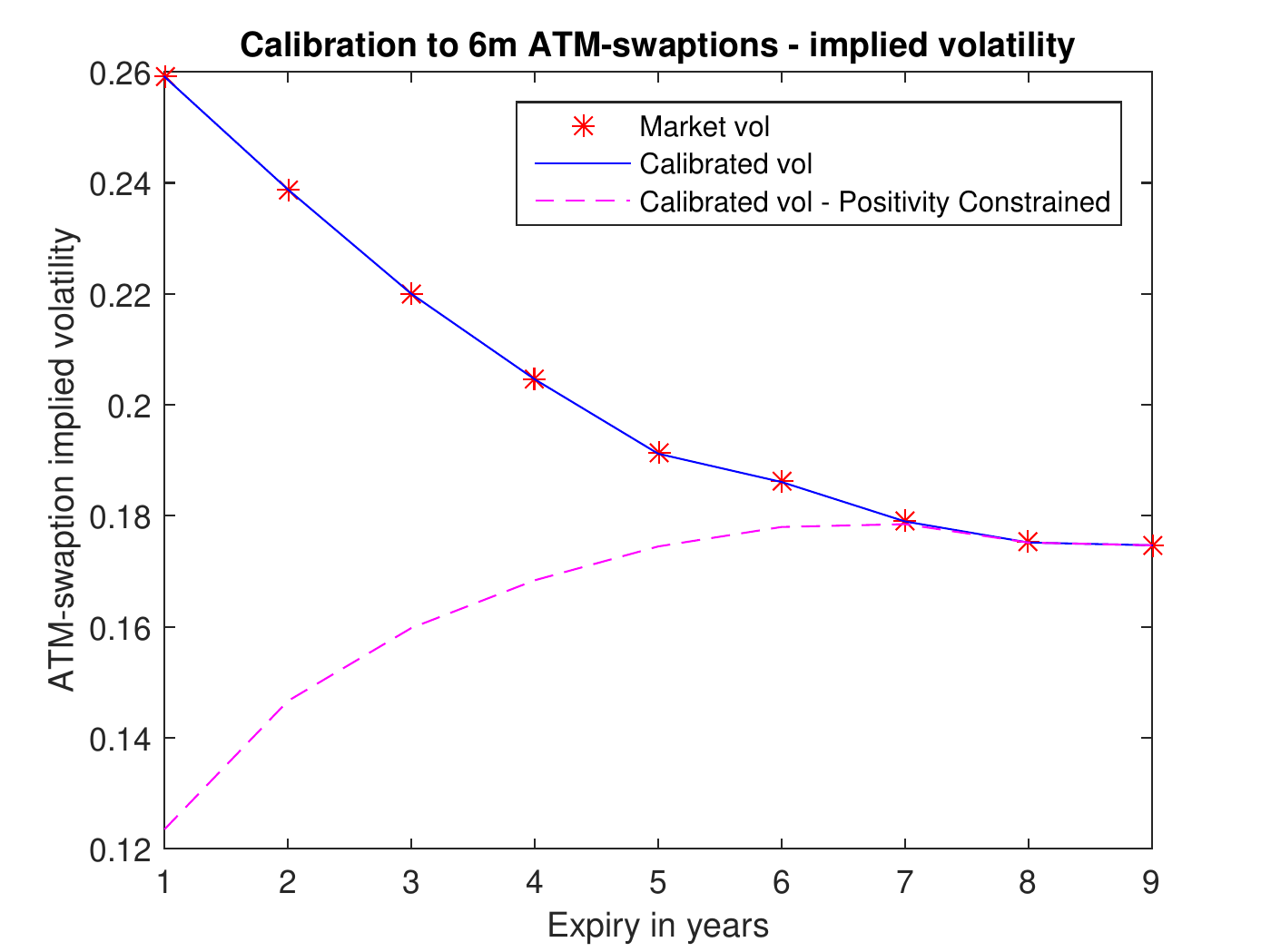}
        \includegraphics[width=.49\textwidth,,height=6cm]{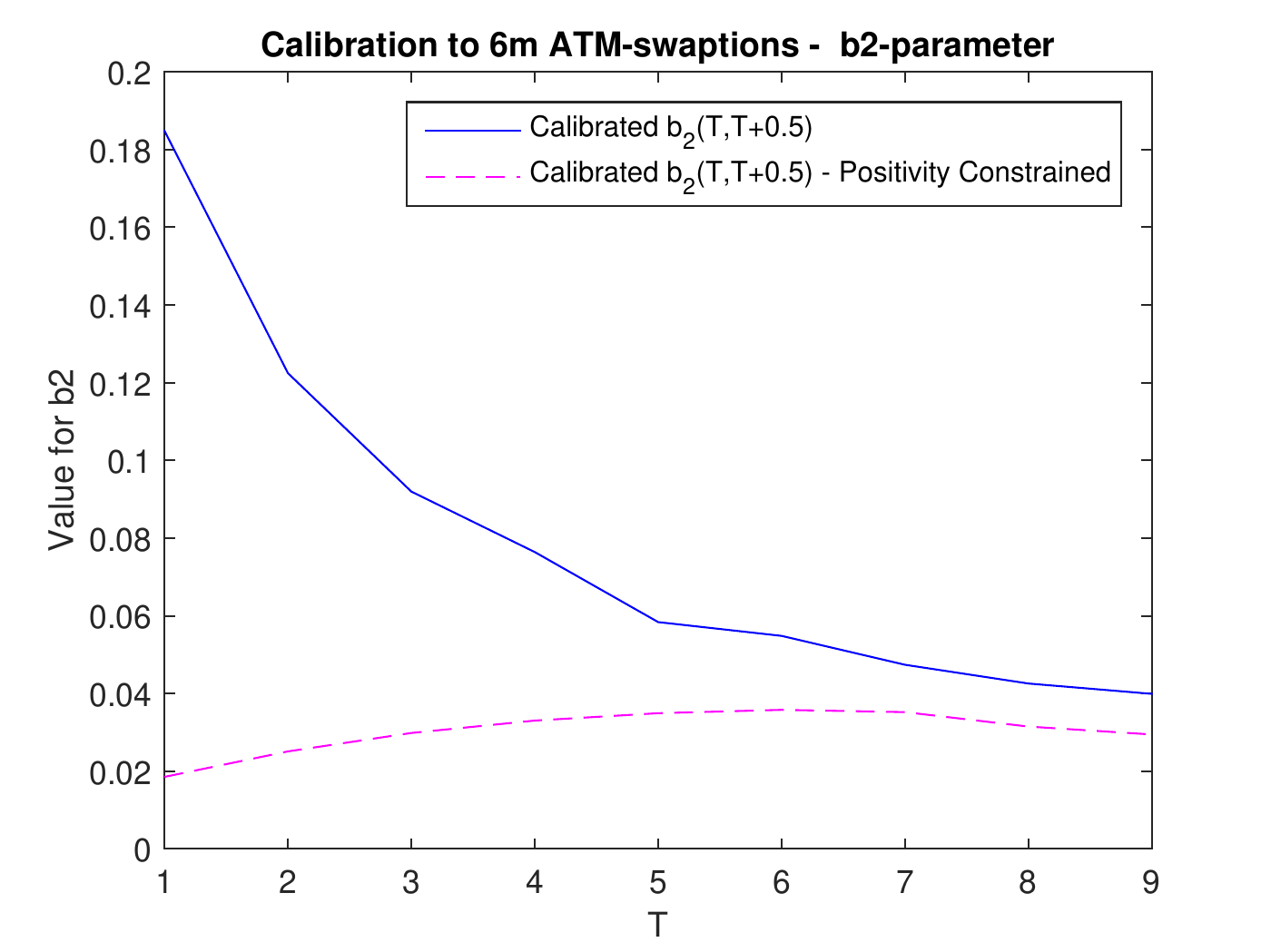}
\caption{Exponential-NIG calibration. {\it (Left)} Fit to ATM swaption implied volatility term structures. {\it (Right)} Calibrated values of the $b_2$ parameters. {\it (Top)} $\delta=3m$. {\it (Bottom)} $\delta=6m$.}\label{f:NIG1dvolsandparms}
\end{figure}

\begin{figure}[htbp]
  \centering
        \includegraphics[width=.49\textwidth,,height=6cm]{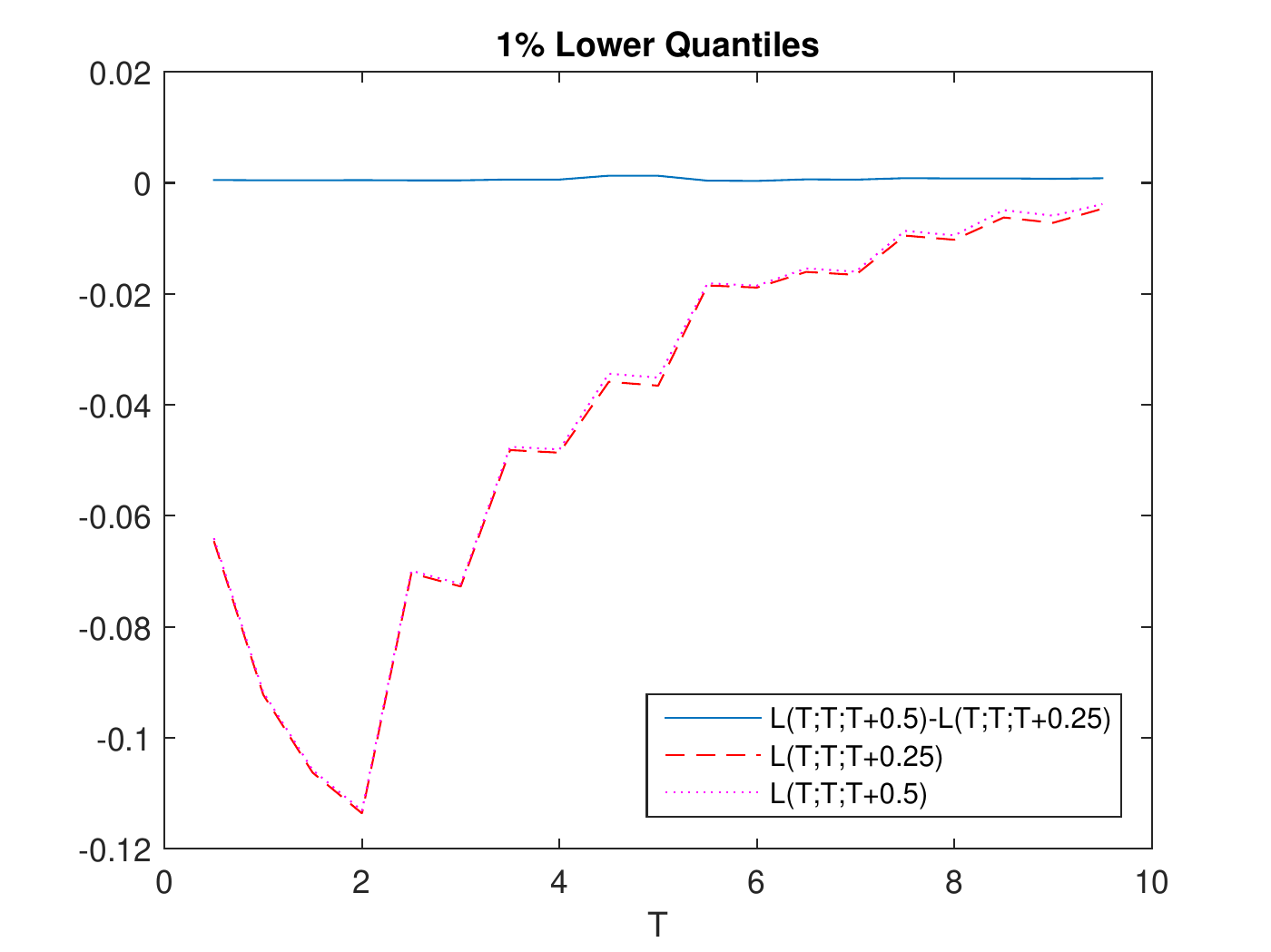}
\caption{Exponential-NIG calibration calibration. 1\% lower quantiles.}\label{f:NIG1dQuantiles}
\end{figure}

The one-factor model, which is driven by a Gaussian factor $\{A_t^{(2)}\}$, is able to capture the level of the volatility smile. Nevertheless, the model implied skew is slightly different from the market skew. To overcome this issue, we now consider a one-factor model driven by a richer family of L\'evy processes. The process $\{A^{(2)}_t\}$ is now assumed to be the exponential normal inverse Gaussian (NIG) $\M$-martingale
\begin{equation}\label{nig}
A^{(2)}_t = \exp\left(\theX ^{(2)}_t - t\psi(1)\right)-1,  
\end{equation}
where $\{X ^{(2)}_t\}$ is an $\M$-NIG-process with cumulant $\psi (z)$, see \eqref{e:cum},  
expressed in terms of the  parametrisation\footnote{The \citeN{Barndorff-Nielsen97} parametrisation is recovered by setting $\mu=0$, $\alpha=\Frac 1\sigma\sqrt{\Frac{\theta_i^2}{\sigma_i^2}+\nu_i^2},\;\beta=\Frac{\theta_i}{\sigma_i^2}$ and $\delta=\sigma\nu$.} $(\nu  , \theta ,\sigma )$ from \cite{ContTankov03} as
\begin{align}
 \label{eq:NIG-cum}
\psi (z)=-\nu \left(\sqrt{\nu^2 - 2 z\theta - z^2 \sigma^2} -\nu\right),\ 
\end{align}
where $\nu , \sigma  > 0$ and $\theta \in \R.$ The parameters that need to be calibrated at first are $\nu, \theta, \sigma$ and $b = b_2(9,9.25)=b_2(9.25,9.5)=b_2(9.5,9.75) =b_2(9.75,10)$. After the calibration, we obtain
\begin{equation*}
b =    0.0431 , \nu =   0.2498, \theta=  -0.0242  ,\sigma = 0.1584 .
\end{equation*}
Imposing $b\leq L(0;9.75,10)= 0.0328$ to get positive rates we obtain instead
\begin{equation*}
b =     0.0291  , \nu =   0.1354 , \theta=  -0.0802  ,\sigma = 0.3048 .
\end{equation*}
The two fits are plotted in Figure \ref{f:NIG1dSmile}. Here, imposing positivity comes at a much smaller cost when compared to the one factor Gaussian case. The NIG process has a richer structure (more parametric freedom) and therefore is able to compensate for an imposed smaller level of the parameter $b_2$. 
We continue with the second part of the calibration of which results are found in Figure \ref{f:NIG1dvolsandparms}. Here we see that enforcing positivity may have a small effect on the smile but it means that the volatility structure cannot be made to match swaptions with maturity smaller than 7 years. Thus, enforcing positivity in this model produces limitations which we wish to avoid. 
In Figure \ref{f:NIG1dQuantiles}, we plot lower quantiles for the rates and spreads as for the one-factor lognormal model. While spot spreads remain positive, the levels do not, and, as shown, the model assigns an unrealistically high probability mass to negative values. In fact the model assigns a 1\% probability to rates falling below -12\% within 2 years! Thus, the one-factor exponential-NIG model loses much of its appeal for it cannot, in a realistic manner, be made to fit long-term smiles and shorter-term ATM volatilities.
%%%%%%%%%%%%%%%%%%%%%%%%%%%%%%%%%%%%%%%%%%%%%%%%%%%%%%%%%%%
\subsection{Calibration of a two-factor lognormal model}\label{s:calln2D}
\begin{figure}[htbp] 
  \centering
 \includegraphics[width=.49\textwidth,,height=6cm]{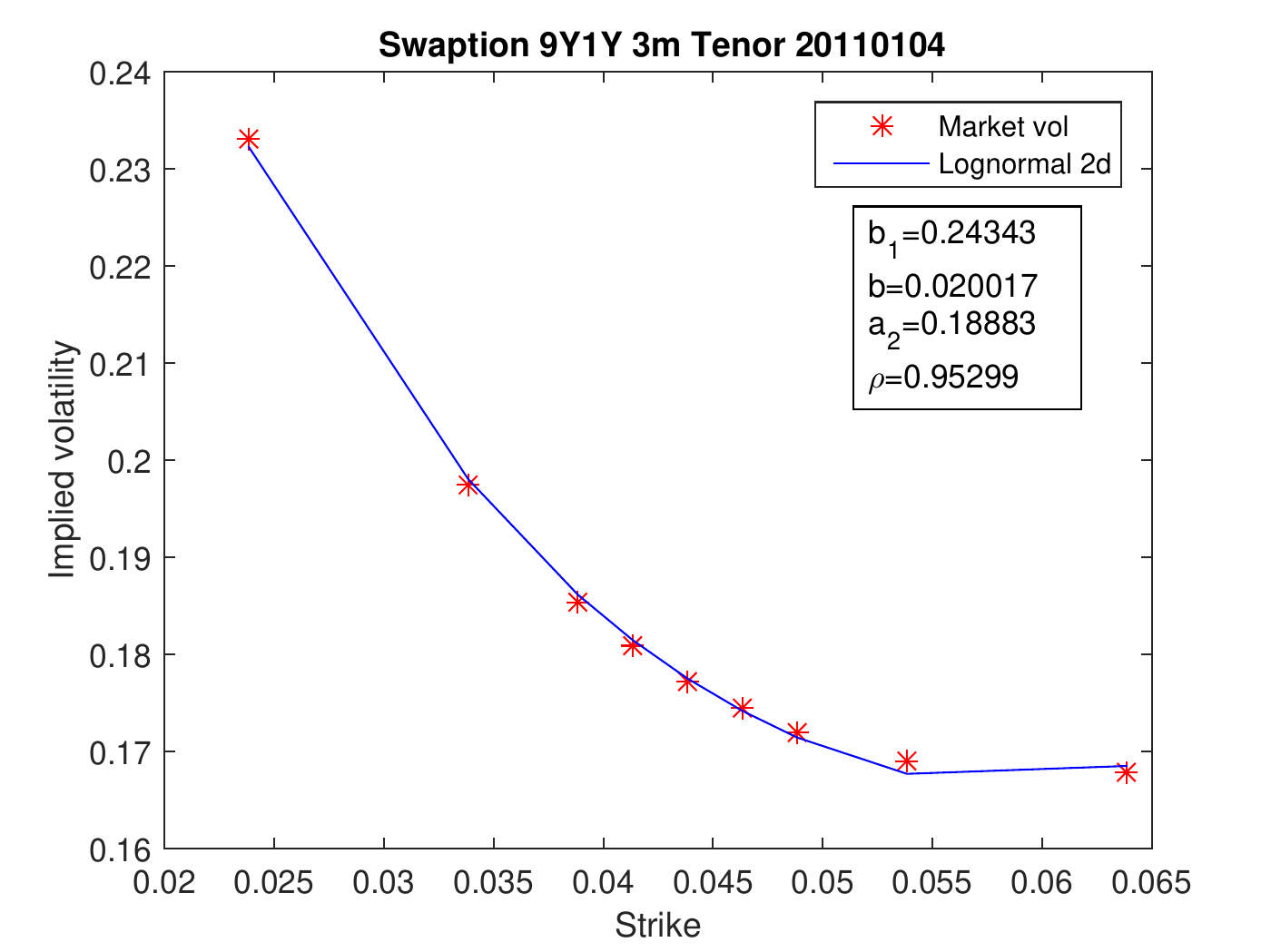} \includegraphics[width=.49\textwidth,,height=6cm]{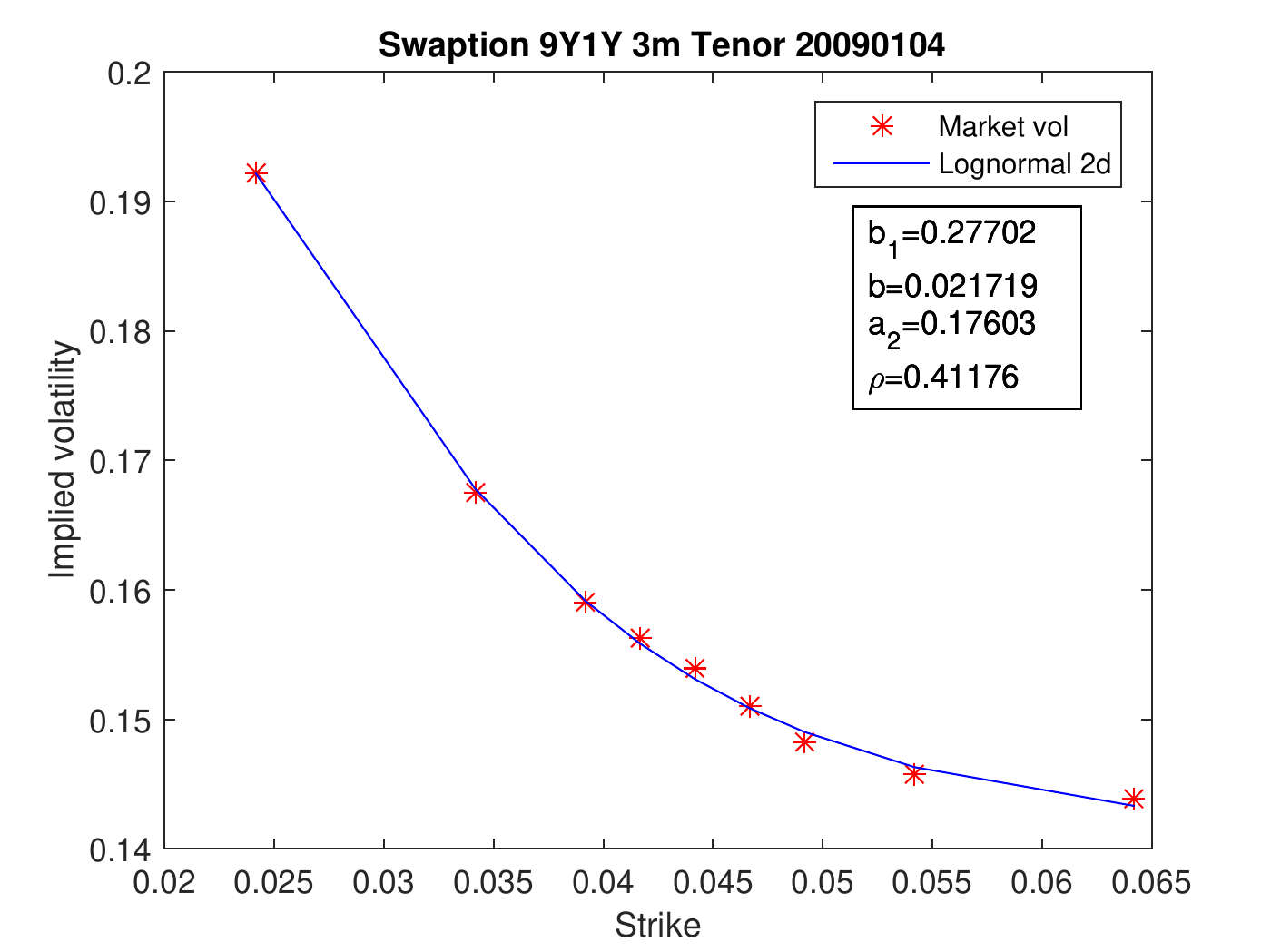}\\
 \includegraphics[width=.49\textwidth,,height=6cm]{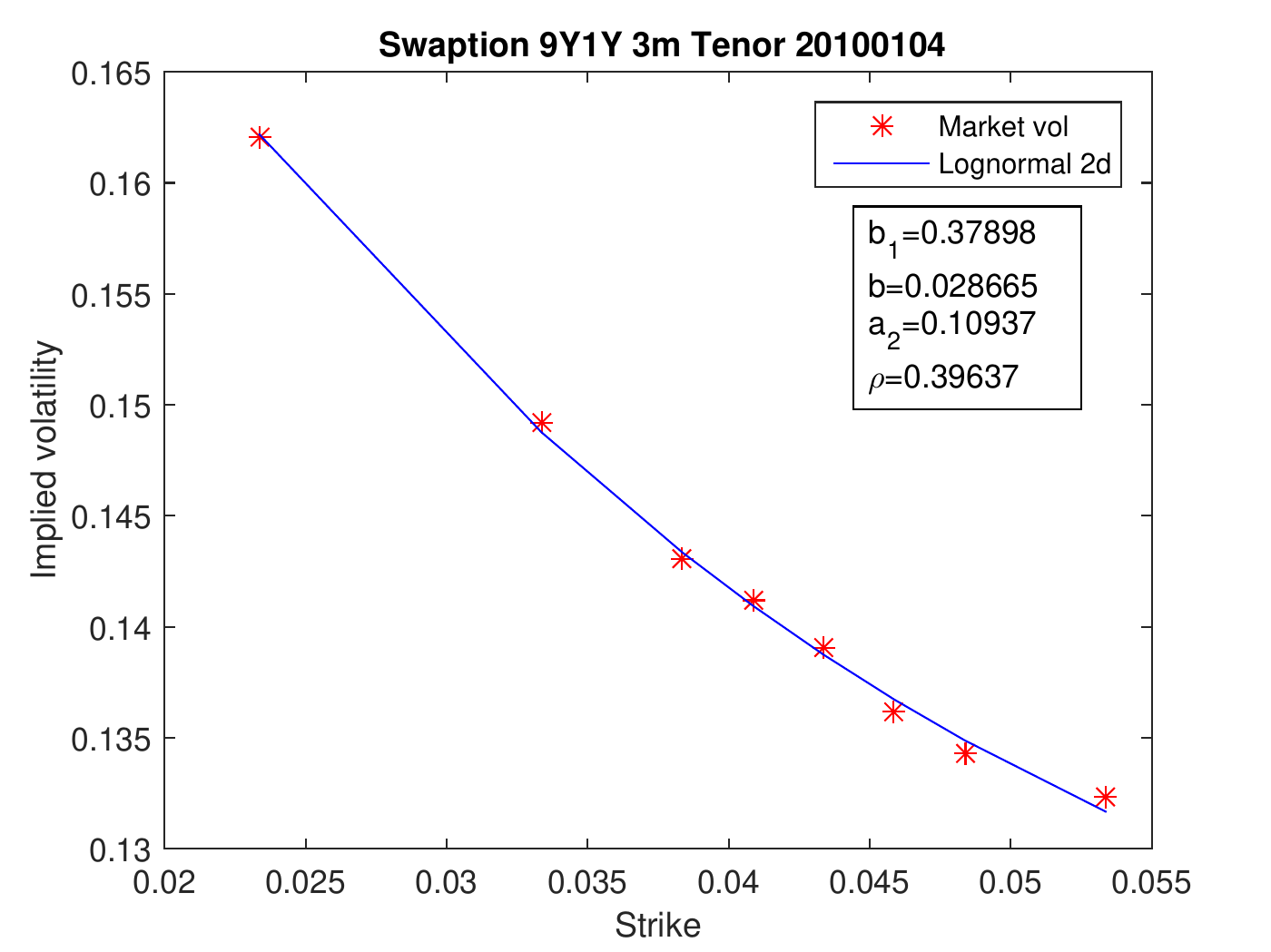} \includegraphics[width=.49\textwidth,,height=6cm]{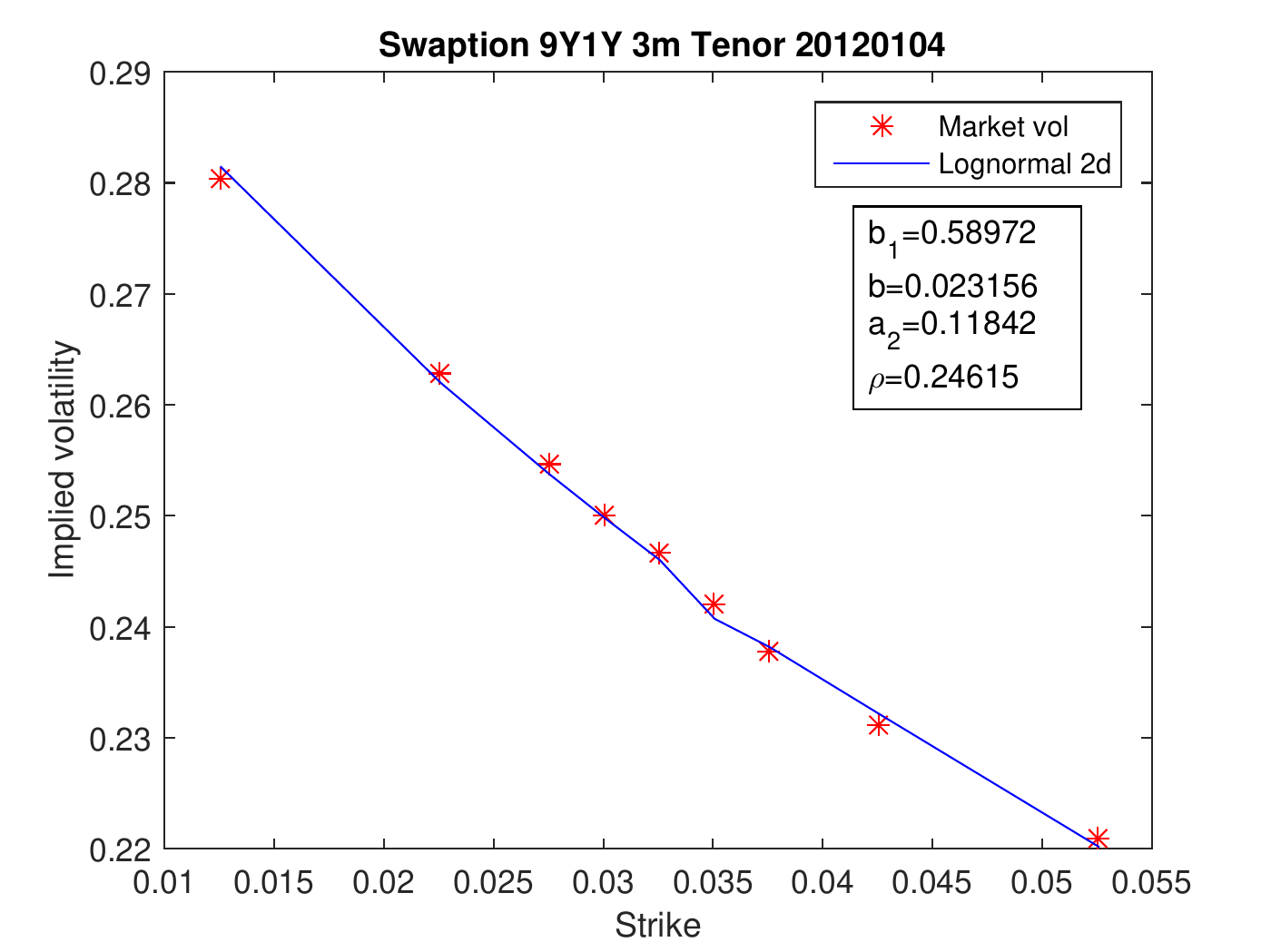}\\
\caption{Lognormal two-factor calibration.}\label{f:Gauss2dSmile}
\end{figure}

\begin{figure}[H]
  \center
        \includegraphics[width=.49\textwidth,,height=6cm]{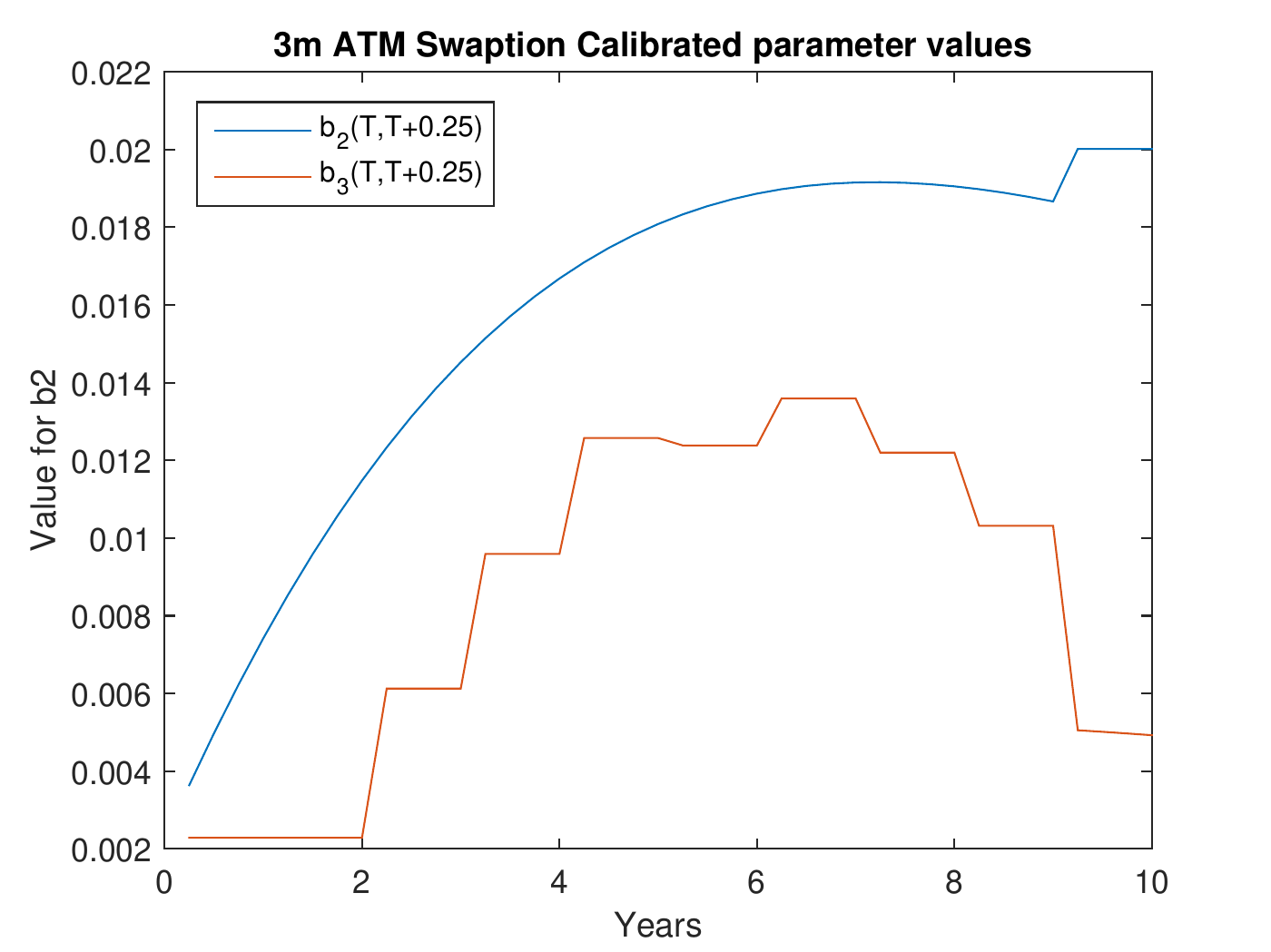}
        \includegraphics[width=.49\textwidth,,height=6cm]{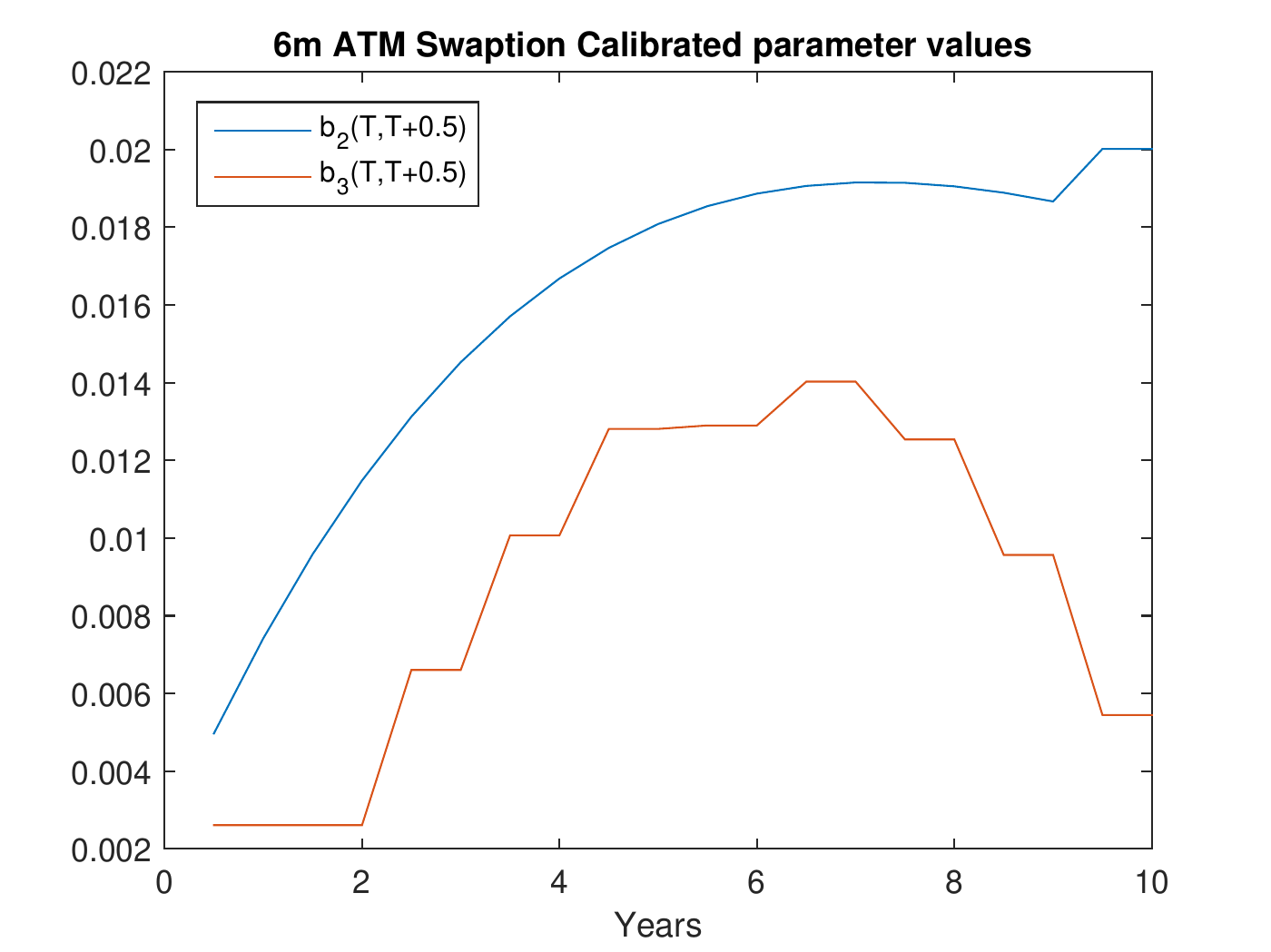}
\caption{Two-factor lognormal calibration. {\it (Left)} Parameter values fitted  to three-month ATM swaption implied volatility term structures. {\it (Right)}  Parameter values fitted  to six-month ATM swaption implied volatility term structures.} \label{f:Gauss2dATMparms}
\end{figure}

The necessity to produce a better fit to the smile than what can be achieved with the one-factor Gaussian model, while maintaining positive rates and spreads, leads us to proposing the two-factor specification presented in \sr{ss:ln2}.
This model is heavily parametrised and the parameters at hand are not all identified by the considered data. We therefore fix the following parameters:
\begin{align}
&a_1=1,\quad a_3=1.6,\\
&b_3(T,T+0.25) = 0.15 L(0;T;T+0.25),\quad T\in [9,9.75],\\
&b_2(T,T+0.25) = 0.55L(0,T;T+0.25),\quad T\in [0,8.75].
\end{align}
We assume that $b_1$ is constant, i.e. $b_1=b_1(T)$ for $T\in [0,10]$, and that $b_3,$ outside of the region defined above, is piecewise constant such that $b_3(T,T+0.25)=b_3(T+0.25,T+0.5)=b_3(T+0.5,T+0.75)=b_3(T+0.75,T+1)$ for each $T=0,1\dots,8$ and $b_3(T,T+0.5)=b_3(T+0.5,T+1)$ holds for each $T=0,1\dots,9$. We furthermore assume that  $b_2(T,T+0.5)= b_2(T,T+0.25),\quad T\in [0,9.5]$. These somewhat {\it ad hoc} choices are done with a view towards $b_2$ and $b_3$ being fairly smooth functions of time.  We herewith apply a slightly altered procedure to calibrate the remaining parameters if compared to the scheme utilised for the one-factor models.
\begin{enumerate}
\item We first calibrate to the smile of the $9\times1$ years swaption which gives us the parameters $a_2,\rho$, the assumed constant value of $b_1$, and $b_2(9,9.25)$ to $b_2(9.75,10)$ which are assumed equal to a constant $b$. Similar to the exponential-NIG model, we make use of four parameters in total to fit the smile.
\item The remaining $b_2$ parameters are determined {\it a priori}, so what remains is to calibrate the values of $b_3$. The three-month tenor values $b_3(T,T+0.25)$ for $T\in [0,8.75]$ are  calibrated to ATM, co-terminal swaptions starting from the  8$\times$2 years and then continuing backwards to the 1$\times$9 years instruments. For the six-month tenor products, we calibrate $b_3(T,T+0.5)$ for $T\in [0, 9.5]$ starting with 9$\times$1 years and proceed backwards.
\end{enumerate}
These are the values we obtain from the first calibration phase:  $b_1=0.2434,\quad b=0.02, \quad a_2=0.1888,\quad \rho=0.9530$.
The corresponding fit is plotted in the upper left quadrant of Figure \ref{f:Gauss2dSmile}. 
In order to check the robustness of the calibrated fit through time, we also calibrate to three alternative dates. The quality of the fit appears quite satisfactory and comparable to the exponential-NIG model. For all four dates the calibration is done enforcing the  positivity condition $b_2(T,T+0.25)+b_3(T,T+0.25)\leq L(0;T,T+0.25)$. However, the procedure yields the exact same parameters even if the constraint is relaxed. We thus conclude that a better calibration appears not to be possible for these datasets by allowing negative rates. Note that it is only for our first data set that the calibrated correlation $\rho$ is as high as $0.9530$. In the other three cases we have $\rho=0.4118$, $\rho=0.3964$, and $\rho=0.2461$. Figure \ref{f:Gauss2dATMparms} shows the parameters $b_2$ and $b_3$ obtained at the second phase of the calibration to the data of 4 January 2011. As with the previous model (cf. the left graphs of Figures \ref{f:Gauss1dvolsandparms} and \ref{f:NIG1dvolsandparms}), the volatilities are matched to market data without any error.
We add here that, although not visible from the graphs, the calibrated parameters satisfy the LIBOR spread positivity discussed in Remark \ref{spread-pos}. 

In conclusion, we find that the two-factor log-normal has the ability to fit the swaption smile very well,  it can be controlled to generate positive rates and positive spreads, and it is tractable with numerically-efficient closed-form expressions for the swaption prices. Given these desirable properties, we discard the one-factor models and retain the two-factor log-normal model for all the analyses in the remaining part of the paper.
\section{Basis swap}\label{s:expos}
In this section, we prepare the ground for counterparty-risk analysis, which we shall treat in detail in Section \ref{s:crisk}. A typical multi-curve financial product, i.e. one that significantly manifests the difference between single-curve and a multi-curve discounting, is the so-called basis swap. Such an instrument uconsists of exchanging two streams of floating payments based on a nominal cash amount $N$ or, more generally, a floating leg against another floating leg plus a fixed leg. In the classical single-curve setup, the value of a basis swap (without fixed leg) is zero throughout its life. Since the onset of the financial crisis in 2007, markets quote positive basis swap spreads that have to be added to the smaller tenor leg, which is clear evidence that LIBOR is no longer accepted as an interest rate free of credit or liquidity risk. We consider a basis swap with a duration of ten years where payments based on LIBOR of six-month tenor are exchanged against payments based on LIBOR of three-month tenor plus a fixed spread. The two payment streams start and end at the same times $T_0 =T_0^1=T^2_0$, $T = T^1_{n_1}=T^2_{n_2 }$. The value at time $t$ of the basis swap with spread $K$ is given by
\begin{equation*}
BS_t = N\left(\sum_{i=1}^{n_1} \delta^{6m}_i L(t;T^1_{i-1},T^1_i) - \sum_{j=1}^{n_2} \delta^{3m}_j(L(t;T^2_{j-1},T^2_j) + K\Bd_{t T^2_j}\right)
\end{equation*}
for $t\leq T_0$. After the swap has begun, i.e. for $T_0\leq t<T$, the value is given by
\begin{eqnarray*}
BS_t &=& N\Bigg(\delta^{6m}_{i_t}L(T^1_{i_t-1};T^1_{i_t-1},T^1_{i_t}) + \sum_{i=i_t+1}^{n_1} \delta^{6m}_i L(t;T^1_{i-1},T^1_i) \\
&& -\delta^{3m}_{j_t}\left(L(T^2_{j_t-1};T^2_{j_t-1},T^2_{j_t})  + K\Bd_{t T^2_{j_t}}\right)-  \sum_{j=j_t+1}^{n_2} \delta^{3m}_j\left(L(t;T^2_{j-1},T^2_j) + K\Bd_{t T^2_j}\right)\Bigg),
\end{eqnarray*}
where $T^1_{i_t}$ (respectively $T^2_{j_t}$) denotes the smallest $T^1_i$ (respectively $T^2_i$) that is strictly greater than $t$. The spread $K$ is chosen to be the fair basis swap spread at $T_0$ so that the basis swap has value zero at inception. We have 
\begin{equation*}
K = \dfrac{\sum_{i=1}^{n_1} \delta^{6m}_i L(T_0;T^1_{i-1},T^1_i) - \sum_{j=1}^{n_2} \delta^{3m}_jL(T_0;T^2_{j-1},T^2_j)}{\sum_{j=1}^{n_2} \delta^{3m}_j\Bd_{T_0  T^2_j}}.
\end{equation*}

\begin{figure}[H] 
  \centering
        \includegraphics[width=.49\textwidth,,height=6cm]{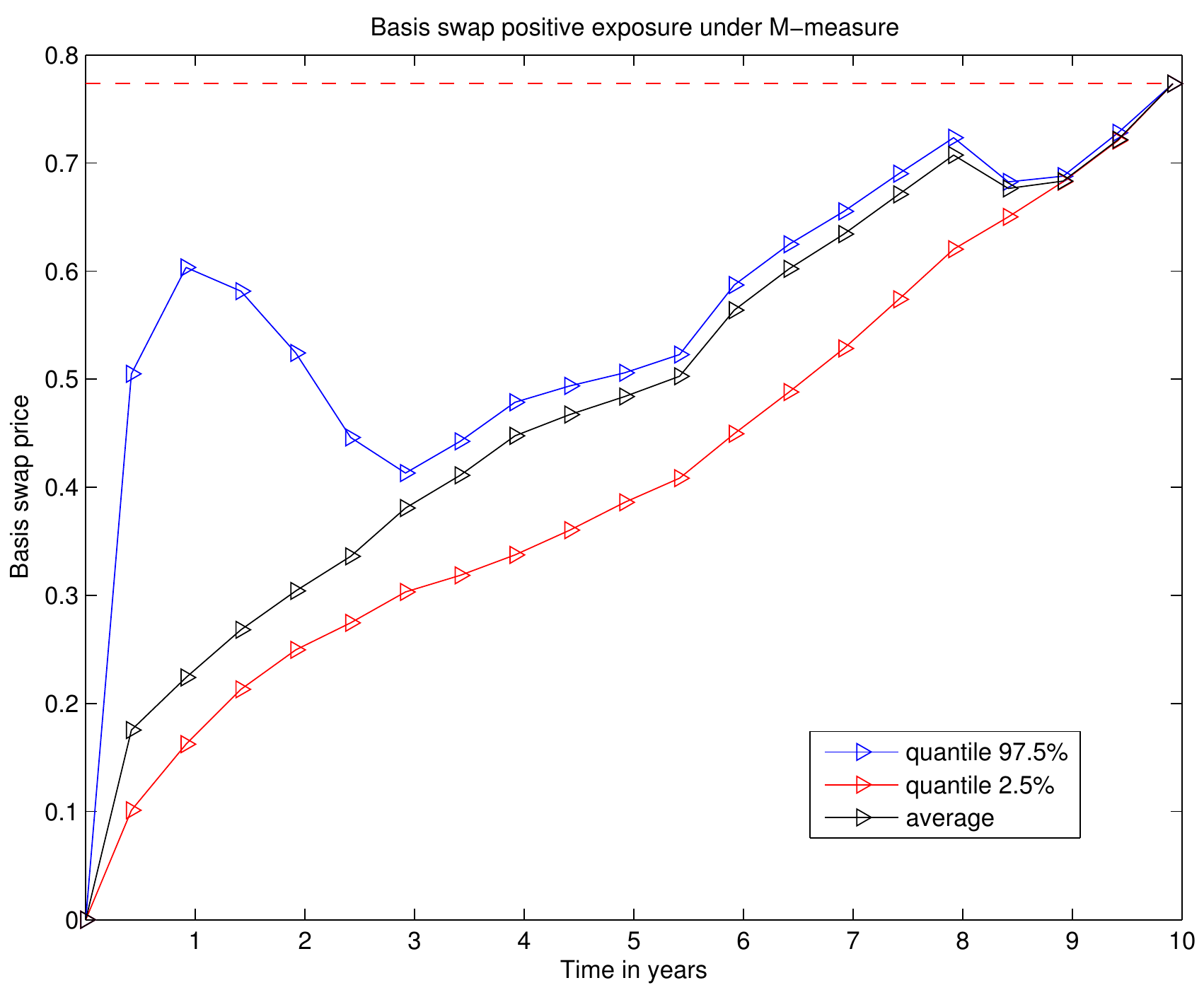}
        \includegraphics[width=.49\textwidth,,height=6cm]{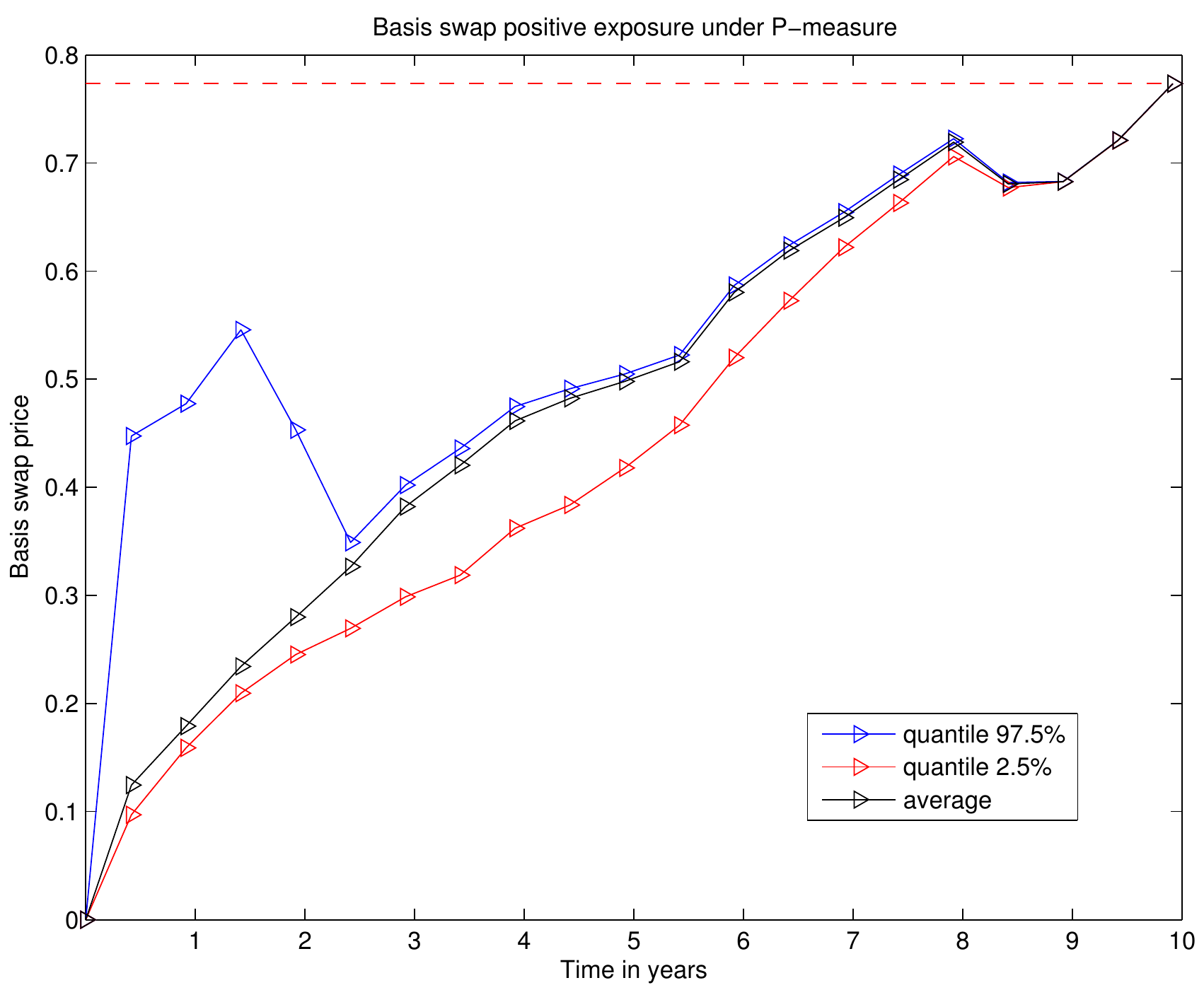}
        \includegraphics[width=.49\textwidth,,height=6cm]{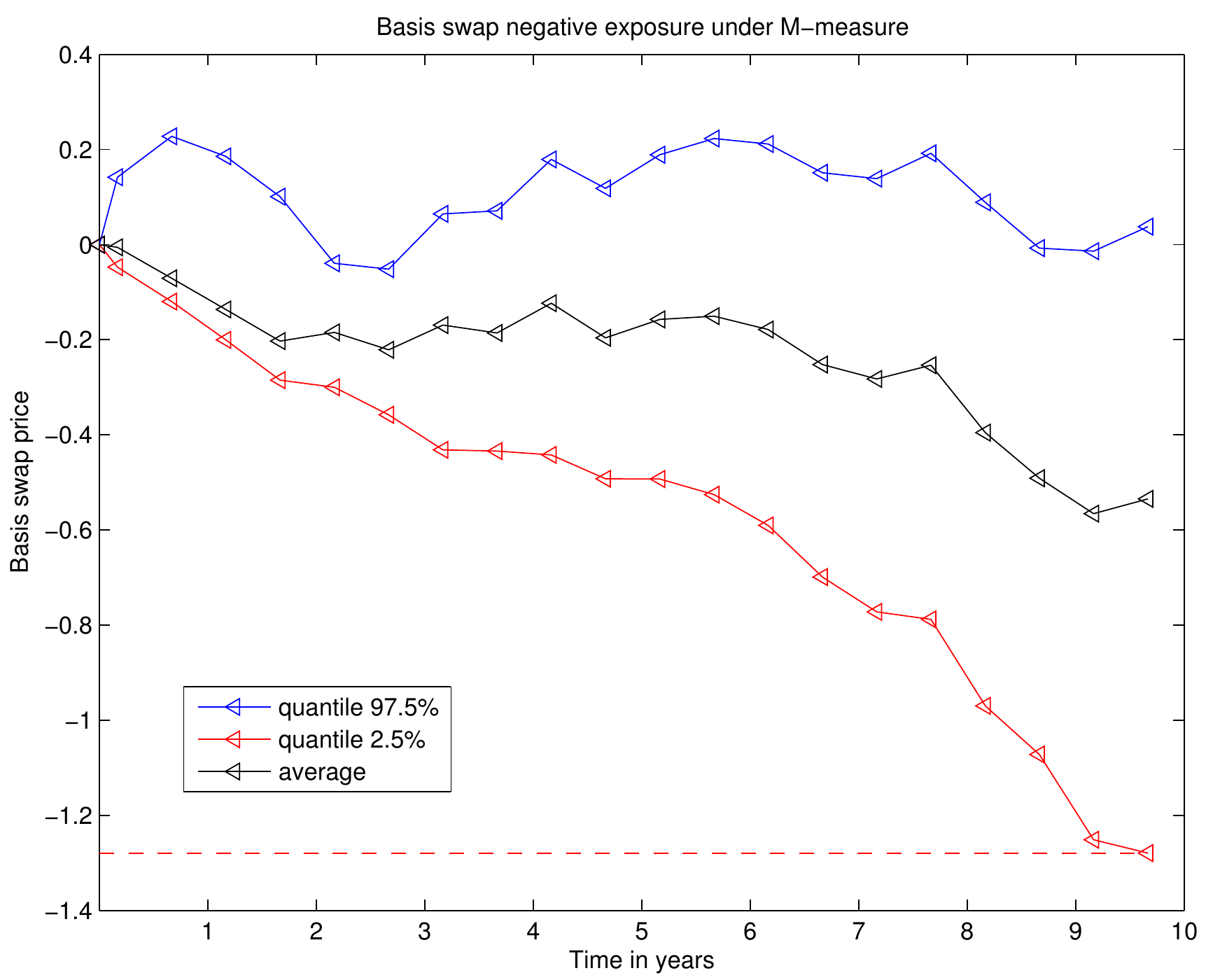}
                 \includegraphics[width=.49\textwidth,,height=6cm]{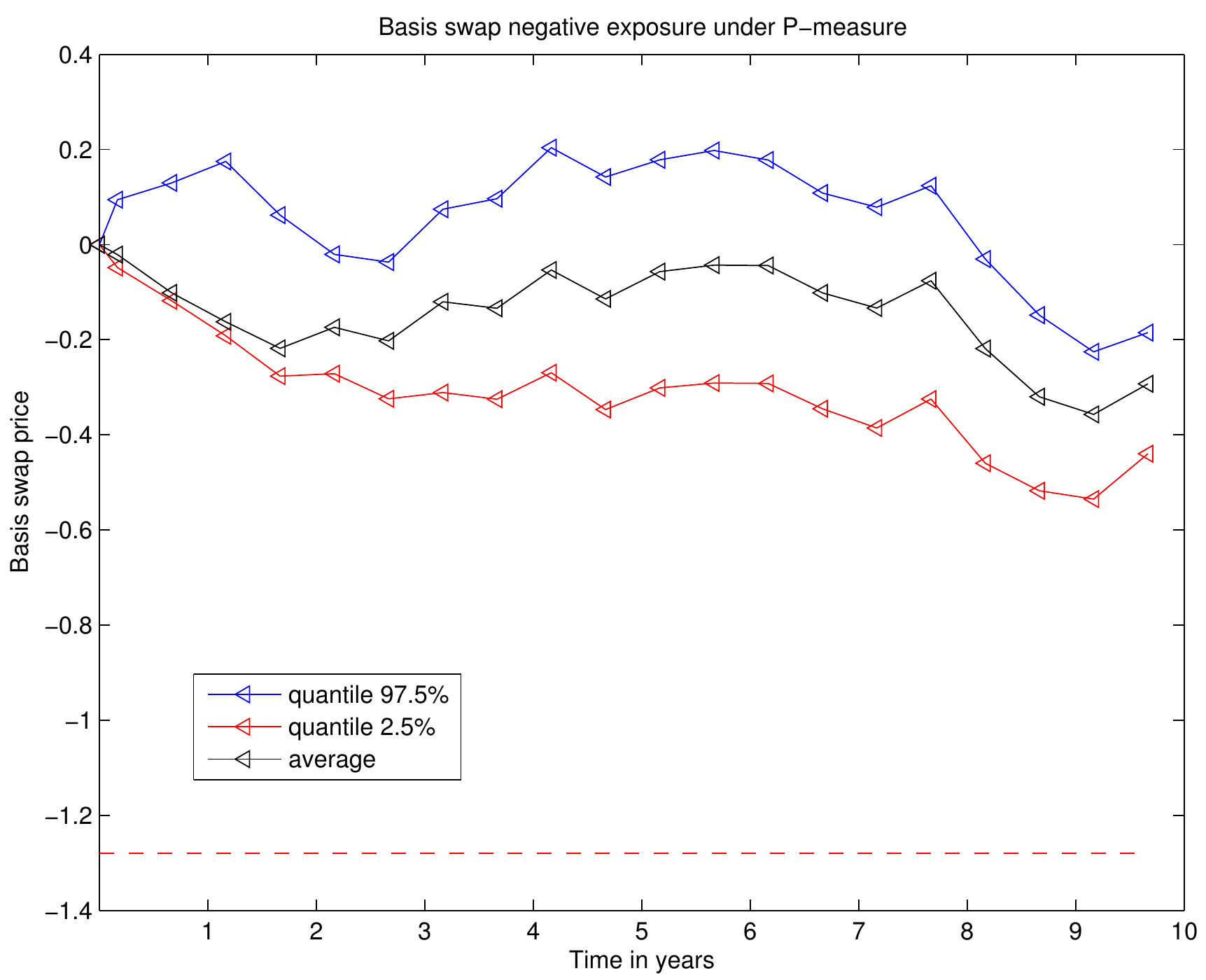}
\caption{Exposures of a basis swap  (price process with mean and quantiles) in the calibrated two-factor Gaussian model. {\it (Top)} Exposure of the basis swap at $t=5m, 11m, etc.$ {\it (Bottom)} Exposure of the basis swap price at $t=2m, 8m, 14m, etc.$ {\it (Left)} Exposure under the $\M$-measure. {\it (Right)} Exposure under the $\P$-measure with the prediction that LIBOR rate $L(10.75 y;10.75 y, 11 y)$ will be either $2\,\%$ with probability $p=0.7$ or $5\,\%$ with probability $1-p=0.3$.}
\label{f7}
\end{figure}

The price processes on which the numerical illustration in Figure \ref{f7} have been obtained was simulated by applying the calibrated two-factor lognormal model developed in Section \ref{s:calln2D}. The basis swap is assumed to have a notional cash amount $N=100$ and maturity $T=10$ years. In the two-factor lognormal setup, the basis swap spread at time $t=0$ is $K=12$ basis points, which is added to the three-month leg so that the basis swap is incepted at par. The $t=0$ value of both legs is then equal to EUR 27.96. The resulting 
risk exposure, in the sense of the expectation and quantiles of the corresponding price process at each point in time, is shown in the left graphs of Figure \ref{f7}, where the right plots correspond to the $\mathbb{P}$ exposure discussed in \sr{s:lrb}. Due to the discrete coupon payments, there are two distinct patterns of the price process {exposure}, most clearly visible at times preceding payments of the six-month tenor coupons for the first one and at times preceding payments of the three-month tenor coupons without the payments of the six-month tenor coupons for the second one. We show the exposures at such respective dates on the upper and lower plots in Figure \ref{f7}. 
%%%%%%%%%%%%%%%%%%%%%%%%%%%%%%%%%%%%%%%%%
\subsection{L\'evy random bridges}\label{s:lrb}
The basis swap exposures in Figure \ref{f7} are computed under the auxiliary $\M$-measure. The XVAs that are computed in later sections are derived from these $\M$-exposures. However, exposures are also needed for risk management and as such need to be evaluated under the real-world measure $\PR$. This means that a measure change from $\M$ to $\PR$ needs to be defined, which requires some thoughts as to what features of a price dynamics under $\PR$ one might like to capture through a specific type of measure change and hence by the induced $\PR$-model. In other words, we design a measure change so as to induce a particular stochastic behaviour of the $\{A_t\}$ processes under $\PR$, and in particular of the underlying Markov processes $\{X_t\}$ driving them. 
 
A special case we consider in what follows is where $\{X_t\}$ is a L\'evy process under $\M,$ while it adopts the law of a corresponding (possibly multivariate, componentwise) L\'evy random bridge (LRB) under $\PR$. Several explicit asset price models driven by LRBs have been developed in \citeN{Mac}. The LRB-driven rational pricing models have a finite time horizon. The LRB is characterised, apart from the type of underlying L\'evy process, by the terminal $\PR$-marginal distribution to which it is pinned at a fixed time horizon $U$. The terminal distribution can be arbitrarily chosen, but its specification influences the behaviour of the LRB as time approaches $U$. In turn, the properties of a specified LRB influence the behaviour of $\{A_t\}$ and hence the dynamics of the considered price process. We see an advantage in having the freedom of specifying the $\PR$-distribution of the factor process at some fixed future date. This way, we can implement experts' opinions (e.g. personal beliefs based on some expert analysis) in the $\PR$-dynamics of the price process as to what level, say, an interest rate (e.g. OIS, LIBOR) is likely to be centred around at a fixed future date. 

The recipe for the construction of an LRB can be found in \cite{hhm1}, Definition 3.1, which is extended for the development of a multivariate LRB in \cite{Mac}. LRBs have the property, as shown in Proposition 3.7 of \cite{hhm1}, that there exists a measure change to an auxiliary measure with respect to which the LRB has the law of the constituting L\'evy process. That is, we suppose the auxiliary measure is $\M$ and we have an LRB $\{X_t\}_{0\le t\le U}$ defined on the finite time interval $[0,U]$ where $U$ is fixed. Under $\M$ and on $[0,U)$, $\{X_t\}$ has the law of the underlying L\'evy process. To illustrate further, let us assume a univariate LRB; the analogous measure change for multivariate LRBs is given in \cite{Mac}. Under $\PR$, which stands in relation with $\M$ via the measure change
\begin{equation}\label{MP}
\eta_t=\frac{\textrm{d}\PR}{\textrm{d}\M}\Big\vert_{\Fc_t}=\int_{\R}\frac{f_{U-t}(z-X_t)}{f_U(z)}\,\nu(\textrm{d}z)\sp t<U,
\end{equation} 
where $f_t(x)$ is the density function of the underlying L\'evy process for all $t\in(0,U]$ and $\nu$ is the $\PR$-marginal law of the LRB at the terminal date $U$, the process $\{X_t\}$ is an LRB (note that the change of measure is singular at $U$). 

Now, returning to the calibrated two-factor lognormal model of \sr{s:calln2D}, but similarly also to the other models in \sr{s:cal}, we may model the drivers $\{X_t^{(1)}\}=\{X_t^{(3)}\}$  and $\{X_t^{(2)}\}$ by \s{two} dependent Brownian random bridges under $\PR$. The computed $\M$ exposures in Figure \ref{f7} thus need to be re-weighted by the corresponding amount $\eta_t$ in order to obtain the $\PR$-exposures of the basis swap. Since here we employ LRBs, we have the opportunity to include an expert opinion through the LRB marginals $\nu$ as to what level one believes the interest rates will tend to by time $U$. The re-weighted $\PR$-exposures of the basis swap are plotted in the graphs of the right-hand side of Figure \ref{f7}. The maximum of the upper quantile curves shown in the graphs is known as the potential future exposure (PFE) at the level $97.5\%$\footnote{In practice, people rather consider the expected positive exposure (expectation of the positive part of the price rather than the price) in the PFE computation, but the methodology is the same.}. 

Hence, we now have the means to propose a risk-neutral model that can be calibrated to option data, and which after an explicit measure change can be applied for risk management purposes while offering a way to incorporate economic views in the dynamic of asset prices. Recalling (\ref{MQ}) and (\ref{MP}), the $\Q$-to-$\PR$ measure change is obtained by
\begin{equation}
\frac{\textrm{d}\PR}{\textrm{d}\Q}\Big\vert_{\Fc_t}=\frac{\eta_t}{\nu_t},
\end{equation}
and the pricing formula for financial assets (\ref{pf-formula}) may be utilised under the various measures as follows:
\begin{eqnarray}
S_{tT} &=&\dfrac{1}{D_t}\E^{\Q}[D_T\,S_{TT}\,\vert\,\Fc_t]=\dfrac{1}{D_t\,\nu_t}\E^{\M}[D_T\,\nu_T\,S_{TT}\,\vert\,\Fc_t]=\dfrac{1}{h_t}\E^{\M}[h_T\,S_{TT}\,\vert\,\Fc_t]\nonumber\\
&=&\dfrac{\eta_t}{D_t\,\nu_t}\E^{\P}\left[\frac{D_T\,\nu_T}{\eta_T}\,S_{TT}\,\Big\vert\,\Fc_t\right]=\dfrac{1}{\pi_t}\E^{\P}[\pi_T\,S_{TT}\,\vert\,\Fc_t],
\end{eqnarray}
for $0\le t\le T<U$ (since we consider price models driven by LRBs). It follows that the pricing kernel is given by $\pi_t=D_t\,\nu_t\,\eta_t^{-1}=\eta_t^{-1}\,h_t$.
Measure changes from a risk-neutral to the real-world probability measure are discussed for similar applications also
elsewhere. For a recent study in this area of research, we refer to, e.g., \cite{hjs}.

%%%%%%%%%%%%%%%%%%%%%%%%%%%%%%%%%%%%%%%%%%%%%%%%%%
\section{Adjustments}\label{s:crisk}
So far we have focused on so-called ``clean computations'', i.e. ignoring counterparty risk and assuming that funding is obtained at the risk-free OIS rate. 
In reality, contractually specified counterparties at the ends of a financial agreement may default, and funding to enter or honour a financial agreement may come at a higher cost than at OIS rate. Thus, various valuation adjustments need to be included in the pricing of a financial position. The price of a counterparty-risky financial contract is computed as the difference between the clean price, as in \sr{s:pg}, and an adjustment accounting for counterparty risk and funding costs.

\subsection{Rational credit model}\label{ss:rcm}
As we shall see below, in addition to their use for the computation of PFE, the exposures in Section \ref{s:expos} can be used to compute various adjustments: CVA (credit valuation adjustment), DVA (debt valuation adjustment) and LVA (liquidity-funding valuation adjustment). With this goal in mind, we equip the bottom-up construction in Section \ref{Bottom_up_Risk_neutral},
the notation of which is used henceforth,
 with a credit component in the following manner. 

We consider $\{X^{(i)}_t\}^{i=1,2,\ldots,n}_{0\le t}$, which are assumed to be $(\{\mathcal{F}_t\},\mathbb{M})$-Markov processes. For any multi-index $(i_1,\ldots,i_d),$ we write
$\mathcal{F}^{(i_1,\ldots,i_d)}_t=\bigvee_{l=1,\ldots, d}\mathcal{F}^{X^{(i_l)}}_t.$ The (market) filtration $\{\mathcal{F}_t\}$ is given by
$\{\mathcal{F}^{(1,\ldots,n)}_t\}$. For the application in the present section, we fix $n=6$. The Markov processes $\{X^{(1)}_t\}=\{X^{(3)}\}$ and $\{X^{(2)}_t\}$ are utilised to drive the OIS and LIBOR models as described in Section \ref{Bottom_up_Risk_neutral}, in particular the zero-initialised  $(\{\mathcal{F}_t\},\mathbb{M})$-martingales $\{A^{(i)}_t\}^{i=1,2,3}$. The Markov processes $\{X^{(i)}_t\}$, $i=4,5,6$, which are assumed  to be $\mathbb{M}$-independent betweeen them and of the Markov processes $i=1,2,3$, are applied to model  $\{\mathcal{F}_t\}$-adapted processes $\{\gamma^{(i)}_t\}^{i=4,5,6}$ defined by
\begin{equation}\label{rg}
\gamma^{(i)}_t =-\Frac{\da (t) + \db (t) \C_{t}}{\a (t) + \b (t) \C_{t}},
\end{equation} 
    where $b_i(t)$ and $c_i(t)$, with $c_i(0)=1$, are non-increasing deterministic functions, and where $\{A^{(i)}_t\}^{i=4,5,6}$ are zero-initialised $(\{\mathcal{F}_t\},\mathbb{M})$-martingales of the form $A(t,X_t^{(i)})$. Comparing with (\ref{rateBU}), we see that (\ref{rg}) is modelled in the same way as the OIS rate (\ref{rateBU}), \s{non-negative in particular, as an intensity should be (see Remark \ref{Rem:pos-intensity})}.  

In line with the ``bottom-up'' construction in Section (\ref{Bottom_up_Risk_neutral}), we now introduce a density $(\{\mathcal{F}_t\},\mathbb{M})$-martingale $\{\mu_t\nu_t\}_{0\le t\le T}$ that induces a measure change from $\mathbb{M}$ to the risk-neutral measure $\mathbb{Q}$:
\bel
&\Frac{d\mathbb{Q}}{d\mathbb{M}} 
 \Big|_{\mathcal{F}_t}=\mu_t \nu_t \qquad (0\leq t\leq T)
\eel
where $\{\mu_t\}$ is defined as in \sr{Bottom_up_Risk_neutral}. Here, we furthermore define $\nu_t=\prod_{i\ge4}\nu^{(i)}_t$ where the processes
\[
\nu^{(i)}_t=\mathcal{E}\left(\int_0^{\cdot}\Frac{\db(t)d\C_t}{\da (t)+ \db (t)\C_{t-} }
 \right)
\]
are assumed to be positive true $(\{\mathcal{F}_t\},\mathbb{M})$-martingales. 
\bl\label{l:indep}
Let $\xi$ denote any non-negative $\mathcal{F}^{(1,2,3)}_T$-measurable random variable and let $\chi=\prod_{j\ge 4}\chi_i$ where, for $j=4,5,6$, $\chi_j$ is $\mathcal{F}^{(j)}_T$-measurable. Then
\beql{rm}&
\ER_ t \left[\xi\,\chi \right] =\ER_ t\left[\xi  \right]\prod_{j\ge 4}\ER_ t\left[ \chi_i  \right] ,
\eeql 
for $\mathbb{R}=\mathbb{M}$ or $\mathbb{Q}$ and for $0\leq t\leq T$.
\el
\proof 
Since $\mathcal{F}^{(4,5,6)}_T$ is independent of $\mathcal{F}^{(1,2,3)}_t$ and of $\xi$, 
 $$ \EM \left[\xi  \,\big|\, \mathcal{F}^{(1,2,3)}_t \vee  \mathcal{F}^{(4,5,6)}_T\right] =\EM \left[\xi  \,\big|\, \mathcal{F}^{(1,2,3)}_t\right].$$ 
Therefore,
 \bel&
\EM_ t \left[\xi\,  \chi  \right] 
=\EM  \left[\EM \left[\xi\,\chi \,\big|\, \mathcal{F}^{(1,2,3)}_t \vee  \mathcal{F}^{(4,5,6)}_T\right] \,\big|\, \mathcal{F}^{(1,2,3)}_t \vee \mathcal{F}^{(4,5,6)}_t \right] \\&\qqq =
\EM  \left[\EM \left[\xi  \,\big|\, \mathcal{F}^{(1,2,3)}_t \vee  \mathcal{F}^{(4,5,6)}_T\right] \chi  \,\big|\, \mathcal{F}^{(1,2,3)}_t \vee \mathcal{F}^{(4,5,6)}_t \right] \\&\qqq=\EM  \left[{\EM \left[\xi  \,\big|\, \mathcal{F}^{(1,2,3)}_t\right]} \chi  \,\big|\, \mathcal{F}^{(1,2,3)}_t \vee \mathcal{F}^{(4,5,6)}_t \right] \\&\qqq=
\EM \left[\xi  \,\big|\,\mathcal{F}^{(1,2,3)}_t\right]\EM  \left[ \chi  \,\big|\, \mathcal{F}^{(1,2,3)}_t \vee \mathcal{F}^{(4,5,6)}_t \right] =\EM_ t\left[\xi   \right]
\EM_ t\left[ \chi  \right].
\eel
Next, the Girsanov formula in combination with the result for $\M$-conditional expectation yields:
\bel&
\EQ_ t \left[\xi \chi  \right]  
= \EM_ t \left[\Frac{\mu_T\nu_T  \xi  \chi  }{\mu_t\nu_t }\right]
= \EM_ t \left[\Frac{ \mu_T \xi }{  \mu_t } \right]
 \EM_ t \left[\Frac{ \nu_T \chi }{ \nu_t } \right]
\\&\qqq
=\EM_ t \left[\Frac{ \nu_T \mu_T \xi }{ \nu_t  \mu_t } \right]
 \EM_ t \left[\Frac{\mu_T \nu_T \chi }{\mu_t \nu_t } \right] 
= \EQ_ t\left[\xi   \right]
\EQ_ t\left[ \chi  \right].\eel
The result remains to be proven for the case $\xi=1$, which is done similarly.\quad\Finproof\\

For the XVA computations, we shall use a reduced-form counterparty risk approach in the spirit of \cite{crepey}, where the default times of a bank ``b'' (we adopt its point of view) and of its counterparty ``c''
are modeled in terms of
three Cox times $\tau_i$ defined by
\beql{Cox-time}
&\tau_i=\inf\left\{t>0\,\big\vert\, \int^t_{0}\gamma^{(i)}_s\,d s \geq {E}_i\right\}
\eeql
Under $\mathbb{Q}$, the random variables $E_i$ ($i=4,5,6$) are independent and exponentially distributed. Furthermore, $\tau_c=\tau_4\wedge\tau_{6}\sp \tau_b=\tau_{5}\wedge\tau_{6}$, hence
$\tau=\tau_b\wedge\tau_c=\tau_4\wedge\tau_{5}\wedge\tau_{6}$. 

We write
\bel
\gamma^c_t=\gamma^{(4)}_t+\gamma^{(6)}_t\sp  \gamma^b_t=\gamma^{(5)}_t+\gamma^{(6)}_t\sp  \gamma_t =\gamma^{(4)}_t+\gamma^{(5)}_t+\gamma^{(6)}_t,
\eel
which are
the so called $(\{\mathcal{F}_t\},\mathbb{Q})$-hazard intensity processes of the $\{\mathcal{G}_t\}$ stopping times $\tau_c,$ $\tau_b$ and $\tau,$ where the full model filtration $\{\mathcal{G}_t\}$ is given as the market filtration $\{\mathcal{F}_t\}$-progressively enlarged by $\tau_c$ and $\tau_b$ (see, e.g., \cite{Bielecki2009}, Chapter 5).
Writing as before $D_t=\exp(-\int_{0}^t r_s\, ds)$, we note that Lemma \ref{lem:p} still holds in the present setup. That is, 
$$
\thep_t  = c_1(t)  + b_1(t)  A_t^{(1)} = D_t\, \mu_t,
$$ 
an $(\{\mathcal{F}_t\},\mathbb{M})$-supermartingale, assumed to be positive (e.g. under an  exponential L\'evy martingale specification for $A^{(1)}$ as of Example \ref{e:theelecont}).
Further, we introduce $Z^{(i)}_t =\exp(-\int_{0}^t \gamma^{(i)}_s\, ds)$,  for $i=4,5,6$, and
obtain analogously that
\begin{equation}\label{k-super}
{\theq^{(i)}_t} :=  {\a(t)+\b(t)\C_t }=Z^{i}_t\, \nu^{(i)}_t.
\end{equation} 
With these observations at hand, the following results follow from Lemma
\ref{l:indep}.
We write $k_t=\prod_{i\ge 4} k^{(i)}$ and $Z_t=\prod_{i\ge 4}Z^{(i)}_t$.

\begin{Proposition}\label{p:form} The identities \qr{e:z} and \qr{LIBOR} still hold in the present setup, that is
\beql{e:first}
&\Bd_{tT}=\EQ_ t\left[\e^{-\int_{t}^T r_s\, ds}\right] 
=\EQ_ t \left[\Frac{D_T}{D_t} \right]
=\EM_ t \left[\Frac{\thep_T  }{\thep_t } \right]=\Frac{\thea (T) + \theb (T)\theC_{t}}{\thea (t) + \theb (t)\theC_{t}}
\eeql
and, for $t\le T_{i-1}$, 
\begin{equation}\label{LIBORbis}
L(t;T_{i-1},T_i)=\dfrac{L(0;T_{i-1},T_i) + b_2(T_{i-1},T_i)A_t^{(2)} + b_3(T_{i-1},T_i)A_t^{(3)}}{\Bd_{0t} + b_1(t)A_t^{(1)}}.
\end{equation}
Likewise,
\beql{e:Zga}
& \EQ_ t \left[\e^{-\int_{t}^T \gamma_s\, ds}\right]  =\EQ_ t \left[\Frac{ Z _T}{ Z _t} \right] 
=\EM_ t \left[\Frac{  \theq _T  }{  \theq _t }  \right] 
=\prod_{i=4,5,6}\Frac{\a (T) + \b (T) \C_{t}}{\a (t) + \b (t) \C_{t}},
\eeql

\begin{eqnarray}\label{e:Zgb}
 \EQ_ t  \left[\e^{-\int_{t}^T \gamma_s\, ds}\,\gamma^c_T\right]  
 &=&{-} \EQ_ t \left[\Frac{Z^{(5)}_T  }{Z^{(5)}_t  } \right]  \partial_T\, \EQ_ t \left[\Frac{Z^{(4)}_T   Z^{(6)}_T}{Z^{(4)}_t   Z^{(6)}_t} \right] \nonumber\\
&=&{-} \EQ_ t \left[\e^{-\int_{t}^T \gamma_s\, ds}\right]\sum_{i=4,6}
\Frac{\da (T) + \db (T) \C_{t}}{\a (T) + \b (T) \C_{t}}, 
\end{eqnarray} 

\beql{e:Zgc}
& \EQ_ t  \left[ \e^{-\int_{t}^T (r_s+\gamma^c_s) ds}\right] =\EQ_ t \left[\Frac{D_T Z^{(4)}_T Z^{(6)}_T}{D_t  Z^{(4)}_t   Z^{(6)}_t } \right]=
\prod_{i=1,4,6}\Frac{\a (T) + \b (T) \C_{t}}{\a (t) + \b (t) \C_{t}}.
\eeql
\end{Proposition}
\proof Using Lemma \ref{l:indep}, we compute
\begin{eqnarray}
\EQ_ t\left[\e^{-\int_{t}^T r_s ds}\right] 
&=&\EQ_ t \left[\Frac{D_T}{D_t} \right]
=\EM_ t \left[\Frac{\thep_T   \nu _T }{\thep_t   \nu _t} \right] 
=\EM_ t \left[\Frac{\thep_T  }{\thep_t } \right] 
\EM_ t \left[ \Frac{  \nu _T }{ \nu _t } \right]\nonumber\\ 
&=&\EM_ t \left[\Frac{\thep_T  }{\thep_t } \right]=\Frac{\thea (T) + \theb (T)\theC_{t}}{\thea (t) + \theb (t)\theC_{t}},
\end{eqnarray}
where the last equality holds by Lemma \ref{lem:p}. This proves \qr{e:first}. The other identities are proven similarly.~\Finproof\\
\begin{Remark}\label{Rem:pos-intensity}\em
Equations (\ref{e:first}) and (\ref{e:Zga}) are similar in nature and appearance. As it is the case for the resulting OIS rate $\{r_t\}$ (\ref{rateBU}), the fact that (\ref{k-super}) is designed to be a supermartingale has as a consequence that the associated intensity (\ref{rg}) is a non-negative process. This is readily seen by observing that $\{\nu^{(i)}_t\}$ is a martingale and thus the drift of the supermartingale (\ref{k-super}) is given by the necessarily non-negative process $\{\gamma^{(i)}_t\}$ that drives $\{Z^{(i)}_t\}$.
\end{Remark}

At time $t=0$, all the $A^{(i)}_0=0,$ hence only the terms $c_i(T)$ remain in these formulas. 
Since the formulas \qr{e:first} and \qr{LIBORbis} are not affected by the inclusion of the credit component in this approach, the valuation of the basis swap of \sr{s:expos} remains unchanged. By making use of the so-called ``Key Lemma'' of credit risk, see for instance \cite{Bielecki2009}, the identity \qr{e:Zgc} is the main building block for the pre-default price process of a ``clean'' CDS on the counterparty (respectively the bank, substituting $\tau_b$ for $\tau_c$ in this formula).  In particular, the identities at $t=0$
\begin{eqnarray}\label{e:jalon}
 \EQ   \left[ \e^{-\int_{0}^T (r_s+\gamma^c_s) ds}\right]&=&\thea (T) c_{4} (T)  c_{6} (T),\\
 \EQ   \left[ \e^{-\int_{0}^T (r_s+\gamma^b_s) ds}\right]&=&\thea (T) c_{5} (T)  c_{6} (T),
\end{eqnarray}
for $T\ge 0,$ can be applied to calibrate the functions $c_{i} (T),$ $i=4,5,6$, to CDS curves of the counterparty and the bank, once the dependence on the respective credit risk factors has been specified. The calibration of the ``noisy'' credit model components $b_{i}(T) A^{(i)}_t,$ $i=4,5,6$, would require CDS option data or views on CDS option volatilities. If the entire model is judged underdetermined, more parsimonious specifications may be obtained by removing the common default component $\tau_{6}$ (just letting $\tau_c=\tau_{4}, \tau_b=\tau_{5}$) and/or restricting oneself to deterministic default intensities by settting some of the stochastic terms equal to zero, i.e. $b_i(T)A_t^{(i)}=0,$ $i=4,5$ and/or $6$ (as is the case for the one-factor interest rate models in Section \ref{s:pg}). The core building blocks of our multi-curve LIBOR model with counterparty-risk are the couterparty-risk kernels $\{\theq^{(i)}_t\}$, $i=4,5,6$, the OIS kernel $\{\thep_t\}$, and the LIBOR kernel given by the numerator of the LIBOR process (\ref{LIBOR}). We may view all kernels as defined under the $\mathbb{M}$-measure, {\it a priori}. The respective kernels under the $\mathbb{P}$-measure, e.g. the pricing kernel $\{\pi_t\}$, are obtained as explained at the end of Section \ref{s:expos}.

\subsection{XVA analysis}

In the above reduced-form counterparty-risk setup following \cite{crepey},
given a contract (or portfolio of contracts) with 
``clean'' price process $\{P_t\}$
and a time horizon $T$, the total valuation adjustment (TVA) process $\{\notTheta_t\}$ accounting for counterparty risk and funding cost, can be modelled as a solution to an
equation of the form
\beql{e:tvagenprel}
&
 \notTheta_t=\EQ_t \left[
\int_t^{\Ts}\exp\left(-\int_t^s (r_u+\gamma_u)du\right)\mygb_s ( \notTheta_s)
 ds \right]\sp\ttt,
\eeql
for some coefficient $\{\mygb_t(\vartheta)\}$. We note that \eqref{e:tvagenprel} is a
backward stochastic differential equation (BSDE) for the TVA process $\{\Theta_t\}$. For accounts on BSDEs and their use in mathematical finance in general and counterparty risk in particular, we refer to, e.g.,  \cite{ElkarouiPengQuenez97},
\shortciteN{BrigoMoriniPallavicini12} and \cite{crepey} or \shortcite[Part III]{CrepeyBieleckiBrigo14}. An analysis in line with \cite{crepey} yields a coefficient of the BSDE \eqref{e:tvagenprel} given, for $\vartheta\in\mathbb{R},$ by:
\bll{fdivbila1}
& \mygb_t (\vartheta)  =
 \,
\underbrace{\gamma^c_t \ (1-\thisRm)
(\Bd_{t}-
 \Gamma_t )^{+} }_{\mbox{{\it CVA coefficient (cva$_t$)}}}
\, - \,
\underbrace{\gamma^b_t
(1-\thisR)(\Bd_{t}- \Gamma_t )^{-}}_{\mbox{{\it DVA coefficient (dva$_t$)}}}
 \\
&\quad\quad\, +\,
\underbrace{\thebm_t \Gp _t- \thenewb _t \Gm _t
+\tilde{\lambda}_t\big( P_t -\vartheta- \Gamma_{t} \big)^+  -
\thelambda_t
\big(P_t -\vartheta - \Gamma_{t} \big)^-}_{\mbox{{\it LVA coefficient (lva$_t (\vartheta)$)}}},
\lel
where:
\begin{itemize}
\item[--] $\thisR$ and $\thisRm$ are the recovery rates of the bank towards the counterparty and vice versa.
\item[--] $\Gamma_t=\Gamma^+_t
-\Gamma^-_t$, where $\{\Gamma^+_t\}$ (resp. $\{\Gamma^-_t\}$)
denotes the value process of the collateral posted by the counterparty to the bank (resp. by the bank to the counterparty), for instance $\Gamma_t=0$ (used henceforth unless otherwise stated) or $\Gamma_t=P_t$.
 \item[--] The processes $\{\thebm_t\}$ and $\{\thenewb_t\}$ are the spreads with respect to the OIS short rate $\{r_t\}$ for the remuneration of the collateral $\{\Gamma^+_t\}$ and $\{\Gamma^-_t\}$ posted by the counterparty and the bank to each other.
 \item[--] The process $\{\thelambda_t\}$ (resp. $\{\tilde{\thelambda}_t\}$) is the liquidity funding (resp. investment) spread of the bank with respect to $\{r_t\}$.
By liquidity funding spreads we mean that these are free of credit risk. In particular,
\beqa\label{e:tiba}\tilde{\thelambda}_t={\thelambdam}_t -\gamma^b_t  (1-\Rf),
\eeqa where
$\{{\thelambdam}_t\}$ is the all-inclusive funding borrowing spread of the bank and
where $\Rf$ stands for a recovery rate of the bank to its unsecured lender (which is assumed risk-free, for simplicity, so that in the case of $\{\thelambda_t\}$ there is no credit risk involved in any case).
\end{itemize}
The data $\{\Gamma_t\}$, $\{\thenewb_t\}$ and $\thebm_t$ are specified in a credit support annex (CSA) contracted between the two parties. We note that
\begin{eqnarray}
\EQ_t \left[
\int_t^{\Ts}\exp\left(-\int_t^s (r_u+\gamma_u)du\right) \mygb_s ( \notTheta_s)
 ds \right]
 &=&\EM_t \left[\int_t^{\Ts}\frac{\mu_s\nu_s D_s  Z_s}
{\mu_s\nu_t D_t  Z_t}\mygb_s ( \notTheta_s)
 ds \right]\nonumber \\ 
 &=&\EM_t \left[\int_t^{\Ts}\frac{h_s k_s}
{h_t  k_t}\mygb_s ( \notTheta_s)
 ds \right]. 
\end{eqnarray}
Hence, by setting
$ \tTheta_t= \thep_t\,\theq_t\,\notTheta_t,$ one obtains the following equivalent 
formulation of \eqref{e:tvagenprel} and \eqref{fdivbila1} under $\mathbb{M}$:
\beql{e:tvagenprelM}
&
 \tTheta_t
=
\EM_t\left[\int_t^{\Ts}  \mygh_s ( \tTheta_s)
 ds \right]
\eeql
for $\ttt$ and where
\bll{fdivbila1M}
& \frac{\mygh_t (\tvartheta)}{h_t k_t} =\mygb_t \big(\frac{\tvartheta}{h_t k_t}\big)=
 \,
 \gamma^c_t \ (1-\thisRm)
(\Bd_{t}-
 \Gamma_t )^{+}   
 - 
 {\gamma^b_t
(1-\thisR)(\Bd_{t}- \Gamma_t )^{-}} 
 \\&\qqq +\,
 \thebm_t \Gp _t- \thenewb _t \Gm _t
+\tilde{\lambda}_t\bigg( P_t -\Frac{\tvartheta}{\thep_t\theq_t}- \Gamma_{t} \bigg)^+  -
\thelambda_t
\bigg(P_t -\Frac{\tvartheta}{\thep_t\theq_t} - \Gamma_{t} \bigg)^- . 
\lel
For the numerical implementations presented in the following section, unless stated otherwise, we set:
\beql{e:parame}&\gamma^b=5\% ,\,\gamma^c=7\% ,\,\gamma=10\% \\&\thisR=
\thisRm=40\%
\\
&\thenewb=\thebm =\thelambda=\tilde{\thelambda}=1.5\%.
\eeql
In the simulation grid one time-step corresponds to one month and $m=10^4$ or $10^5$ scenarios are produced. We recall the comments made after \qr{e:jalon} and note that (i) this is a case where default intensities are assumed deterministic, that is $b_iA^{(i)}=0$ ($i=4,5,6$) and (ii) the counterparty and the bank may default jointly which is reflected by the fact that
$\gamma_t<\gamma_t^b+\gamma_t^c.$
\subsubsection{BSDE-based computations} 
The BSDE \eqref{e:tvagenprelM}-\eqref{fdivbila1M} can be solved numerically by simulation/regression schemes similar to those used for the pricing of American-style options, see \cite{CrepeyGerboudGrbacNgor12}, and \cite{CrepeyGrbacNgorSkovmand13}. 
Since in \qr{e:parame} we have $\thelambda_t=\tilde{\thelambda}_t$, the coefficients of the terms $(P_t -\frac{\tvartheta}{h_t k_t}- \Gamma_{t} )^{\pm}$ coincide in \eqref{fdivbila1M}. This is the case of a ``linear TVA'' where the coefficient $\mygh_t$ depends linearly on $\tvartheta$. The
results emerging from the numerical BSDE scheme for \eqref{fdivbila1M} can thus be verified
by a standard Monte Carlo computation. Table \ref{t:1bis} displays the value of the TVA and its CVA, DVA and LVA components at time zero, where the components are obtained by substituting for $\vartheta,$ in the respective term of \eqref{fdivbila1M}, the TVA process $\tilde{\Theta}_t$ computed by simulation/regression in the first place
(see Section 5.2 in \shortciteN{CrepeyGerboudGrbacNgor12} for the details of this procedure).
The sum of the CVA, DVA and LVA, which in theory equals the TVA, is shown in the sixth column. Therefore, columns two, six and seven yield three different estimates for $\tilde{\Theta}_0={\Theta}_0$. Table \ref{t:1ter} displays the relative differences between these estimates, as well as the Monte Carlo confidence interval in a comparable scale, which is shown in the last column. The TVA repriced by the sum of its components is more accurate than the regressed TVA. This observation is consistent with the better performance of \citeN{Longstaff-Schwartz-98} when compared with \citeN{Tsitsiklis-VanRoy-2000} in the case of American-style option pricing by Monte Carlo methods (see, e.g., Chapter 10 in \citeN{Crepey12}).

\begin{table}[htbp]
 \begin{center}
\begin{tabular}{|r||r|r|r|r|r|r|r||r|}
\hline
m & Regr TVA & CVA & DVA & LVA & Sum & MC TVA \\
\hline
$10^4$ & 0.0447 & 0.0614 & -0.0243 & 0.0067 & 0.0438 & 0.0438   \\
\hline
$10^5$ & 0.0443 & 0.0602 & -0.0234 & 0.0067 & 0.0435 & 0.0435 \\
\hline
\end{tabular}
 \caption{TVA at time zero and its decomposition (all quoted in EUR)
computed by regression for $m=10^4$ or $10^5$ against $X^{(1)}_t$ and $X^{(2)}_t$.
{\it Column 2}: TVA $\Theta_0$. {\it Columns 3 to 5}: CVA, DVA, LVA at time zero repriced individually
by plugging $\tilde{\Theta}_t$ for $\tilde{\vartheta}$ in the respective term of \eqref{fdivbila1M}. {\it Column 6}: Sum of the three components. {\it Column 7}: TVA computed by a standard Monte Carlo scheme.}
 \label{t:1bis}
 \end{center}
 \end{table}
 %------------------------------------------------------------------------------------------------------
\begin{table}[htbp]
 \begin{center}
\begin{tabular}{|r||r|r|r|r|r|r|r|}
\hline
m &  Sum/TVA & TVA/MC & Sum/MC & CI/$\!$/$|$MC$|$ \\
\hline
$10^4$& -2.0114\% & 2.0637\% & 0.0108 \%& 9.7471\%  \\
\hline
$10^5$ & -1.7344 \%& 1.7386 \%& -0.0259\% & 2.9380\%  \\
\hline
\end{tabular}
 \caption{Relative errors of the TVA at time zero corresponding to the results of Table \ref{t:1bis}. ``A/B" represents the relative difference $(A-B)/B$. ``CI/$\!$/$|$MC$|$", in the last column, refers to the half-size of the 95\%-Monte Carlo confidence interval divided by the absolute value of the standard Monte Carlo estimate of the TVA at time zero.}
 \label{t:1ter}
 \end{center}
 \end{table}
In Table \ref{t:1new}, in order to compare alternative CSA specifications, we repeat the above numerical implementation in each of the following four cases, with $\thelambdam_t$ set equal to the constant $4.5\%$ everywhere and all other parameters as in \eqref{e:parame}:
\begin{eqnarray} \label{e:speci}
\begin{array}{llll}
1.&(\Rf,\thisR,\thisRm)= (100,40,40)\% ,\,
&Q=P ,\,
& \Gamma=0 \\
2.&(\Rf,\thisR,\thisRm)= (100,40,40)\% ,\,
&Q=P ,\,
& \Gamma=Q=P\\
3.&(\Rf,\thisR,\thisRm)= (40, 40,40)\% ,\,
&Q=P ,\,
& \Gamma=0\\
4.&(\Rf,\thisR,\thisRm)= (100,100,40)\% ,\,
&Q=P ,\,
& \Gamma=0
\end{array}
\end{eqnarray}
\begin{table}[htbp]
 \begin{center}
\begin{tabular}{|r||r|r|r|r|r|r|}
\hline
Case & Regr TVA & CVA & DVA & LVA & Sum & Sum/TVA \\
\hline
1 &0.0776 & 0.0602 & -0.0234 & 0.0408 & 0.0776 & -0.0464 \% \\
\hline
2 & 0.0095 & 0.0000 & 0.0000 & 0.0092 & 0.0092 & -3.6499\% \\
\hline
3 & 0.0443 & 0.0602 & -0.0234 & 0.0067 & 0.0435 & -1.7344 \%\\
\hline
4 & 0.0964 & 0.0602 & 0.0000 & 0.0376 & 0.0978 & 1.4472\% \\
\hline
\end{tabular}
 \caption{TVA at time zero and its decomposition (all quoted in EUR)
computed by regression for $m=10^5$ against $X^{(1)}_t$ and $X^{(2)}_t$.
{\it Column 2}: TVA $\Theta_0$. {\it Columns 3 to 5}: CVA, DVA and LVA at time zero, repriced individually
by plugging $\tilde{\Theta_t}$ for $\tilde{\vartheta}$ in the respective term
of \eqref{e:tvagenprelM}.
{\it Column 6}: Sum of the three components. {\it Column 7}: Relative difference between the second and the sixth columns.}
 \label{t:1new}
\end{center}
\end{table}
Remembering that the $t=0$ value of both legs of the basis swap is equal to EUR 27.96, the number in Table \ref{t:1new} may seem quite small, but one must also bear in mind that the toy model that is used here doesn't account for any wrong-way risk effect (see \citeN{CrepeySong15}). In fact, the most informative conclusion of the table is the impact of the choice of the parameters on the relative weight of the different XVA components.
\subsubsection{Exposure-based computations} 

Let's restrict attention
to the case of interest rate derivatives with
$\{P_t\}$ adapted with respect to $\{\mathcal{F}^{(1,2,3)}_t\}.$ We introduce $c(s)=\prod_{i\ge 4} c_i(s)$ and the function of time
$$EPE(s):=\EM\left[\thep_s P^+_s\right]=\EQ\left[D_s P^+_s\right]\sp \mbox{resp. }\EM\left[\thep_s P^-_s\right]=\EQ\left[D_s P^-_s\right],$$ called the expected positive exposure, resp. expected negative exposure. For an interest-rate swap, the EPE and ENE correspond to the mark-to-market of swaptions with maturity $s$ written on the swap, which can be recovered analytically if available in a suitable model specification. In general, the EPE/ENE can be retrieved numerically by simulating the exposure. 

In view of \eqref{e:tvagenprelM}-\eqref{fdivbila1M}, by the time $t=0$ forms of \qr{e:Zga} and \eqref{e:Zgb},
the noncollateralised CVA at $t=0$
satisfies (for $R_c\neq 1,$ 
otherwise $CVA_0$=0): 
\bel
&  {\frac{1}{(1-R_c )}}CVA_0 
 = \EM\left[\int^T_{0} h_s k_s  \gamma^c_s  P^+_s  ds\right]= 
\int^T_{0}\EM\left[h_s P^+_s\right]\EM\left[k_s \gamma^c_s\right]ds\\&\qqq
=\int^T_{0}\EM\left[h_s P^+_s\right]\EQ\left[Z_s \gamma ^c_s\right] ds
=-\int^T_{0}  EPE(s)\left(\frac{\dot{c}_{6}(s)}{c_{6}(s)}+\frac{\dot{c}_4(s)}{c_4(s)}\right) c(s) ds .
\eel 
Similarly, for the DVA (for $R_b\neq 1$,  
otherwise $DVA_0=0$) we have:
\bel
& {\frac{1}{(1-R_b )}}DVA_0
=-\int^T_{0} ENE(s)\left(\frac{\dot{c}_{-2}(s)}{c_{-2}(s)}+\frac{\dot{c}_{-1}(s)}{c_{-1}(s)}\right)  c (s)
ds.
\eel
For the basis swap of Section ~\ref{s:expos} and the counterparty risk data \eqref{e:parame}, 
we obtain by this manner $CVA_{0}=0.0600$ and $DVA_{0}=-0.0234$,
quite consistent with the corresponding entries of the second row (i.e. for $m=10^5$) in Table \ref{t:1bis}. 
As for the LVA, to simplify its computation, one may be tempted to neglect the nonlinearity that is inherent to $lva_t (\vartheta)$ (unless $\tilde{\lambda}_t=\thelambda_t$), replacing $\vartheta$ by 0 in $lva_t (\vartheta).$ Then, assuming $lva_t(0)\in  \mathcal{X}^{(1,2,3)}_t,$ by \eqref{e:tvagenprel}-\eqref{fdivbila1}, one can compute a linearised LVA at time zero given by
\bel
&\widehat{LVA}_{0} 
 = \EM\left[\int^T_{0}  h_s k_s\, lva_s(0)  ds \right]
 =     \int^T_{0} \EM \left[ h_s\,  lva_s(0)\right] \EM \left[ k_s \right]  ds 
  =\int^T_{0} \EM \left[ h_s\,  lva_s(0)\right]  c (s) ds,
\eel
by \eqref{e:Zga} for $t=0.$
This is based on the expected (linearised) liquidity exposure 
$$\EM\left[\thep_s lva_s(0)\right]=\EQ\left[D_s lva_s(0)\right].$$ 
In case of no collateralisation ($\Gamma_t = 0$) and of deterministic $\tilde{\lambda}_t$ and ${\lambda}_t$, we have
\begin{equation*}
lva_s(0) = \tilde{\lambda}_sP_s^{+} - \lambda_sP_s^{-}\sp
\widehat{LVA}_{0}=\int^T_{0}\left( \tilde{\lambda}_s  EPE(s) - \lambda_s ENE(s)\right)c (s)ds.
\end{equation*}
In case of continuous collateralisation ($\Gamma_t = P_t$) and of deterministic $\bar{b}_t$ and $b_t$, the formulas read
\begin{equation*}
lva_s(0) = \bar{b}_sP^{+}_s - b_sP^{-}_s\sp
\widehat{LVA}_{0}=\int^T_{0}\left( \bar{b}_s  EPE(s) - b_s ENE(s)\right)c (s)ds.
\end{equation*}
As for CVA/DVA, the LVA exposure is controlled by the EPE/ENE functions, but for different ``weighting functions'', depending on the CSA. For instance, 
for the data \eqref{e:parame},
the LVA on the basis swap  of Section~\ref{s:expos} (collateralised or not, since in this case $\bar{b}_t=b_t=\tilde{\lambda}_t=\lambda_t=1.5\%$), we obtain $\widehat{LVA}_{0} =0.0098,$
quite different in relative terms (but these are small numbers) from  the exact (as opposed to linearised) value of 0.0067 in Table \ref{t:1bis}. We note that the stochasticity of the default intensities $\{\gamma_t^{(i)}\}$ is averaged out in all these $t=0$ pricing formulas,
but it would appear in more general $t$ pricing formulas or in the XVA Greeks even for $t=0$.
\\

\noindent{\bf \Large Acknowledgments}
\\

The authors thank M. A. Crisafi, C. Cuchiero, C. A. Garcia Trillos and Y. Jiang for useful discussions, and participants of the first Financial Mathematics Team Challenge, University of Cape Town (July 2014), the 5th International Conference of Mathematics in Finance, Kruger Park, South Africa (August 2014), and of the London-Paris Bachelier Workshop, Paris, France (September 2014) for helpful comments. The research of S. Cr\'epey benefited from the support of the ``Chair Markets in Transition'' under the aegis of Louis Bachelier Laboratory, a joint initiative of \'Ecole Polytechnique, Universit\'e d'\'Evry Val d'Essonne and F\'ed\'eration Bancaire Fran\c caise.

\bibliographystyle{chicago}

\end{document}